\documentclass{svjour3}                     
\smartqed  
\usepackage{graphicx}
\usepackage{natbib}
\usepackage{lineno}
\begin{document}

\title{Trends of continental, zonal and seasonal land temperatures in the 20th century
}

\titlerunning{Trends of temperatures in the 20th century}        

\author{Jouni Takalo        \and
        Kalevi Mursula 
}


\institute{Space physics and astronomy research unit, University of Oulu, POB 3000, FIN-90014, Oulu, Finland
							\email{jouni.j.takalo@oulu.fi}}


\begin{center}
\textbf{\huge Trends of continental, zonal and seasonal land temperatures in the 20th century}
\newline
\newline

\textbf{Jouni Takalo and Kalevi Mursula} 
\newline

Space physics and astronomy research unit, University of Oulu, POB 3000, FIN-90014, Oulu, Finland,
							email: jouni.j.takalo@oulu.fi

\end{center}

\begin{abstract}
We study the evolution of continental, zonal and seasonal land temperature anomalies especially in the early 20th century warming (ETCW) period, using principal component analysis (PCA) and reverse arrangement trend analysis. ETCW is significant in all other continents except for Oceania. Warming in South America is significant from the ETCW onwards, but significant recent warming started in North America and Europe only around 1990. The continental PC2 is related to North Atlantic Oscillation and mainly shows short period differences between North America and Eurasia. The continental PC3 component including the ETCW depicts a 60-70-year oscillation, which is related to Atlantic multidecadal oscillation (AMO). 

The zonal and seasonal PC2s are both correlated with AMO index, but zonal PC3 is related to Southern oscillation index (SOI) and seasonal PC3 best correlated with wintertime El Ni{\~n}o (NINO34 DJF index). In the southern hemisphere, the recent warming starts first closest to the equator in the 1950s and latest in the southernmost zone in the late 1970s. In the two lowest northern zones (EQ-N24, N24-N44) the warming is significant since the ETCW, and increased warming starts in 1970s,  but in two northernmost zones (N44-N64, N64-N90) the cooling after the ETCW delays the start of recent warming until around 1990.
All seasons of the northern hemisphere but no season in the southern hemisphere depict a significant ETCW. Significant recent warming starts in the southern seasons already around 1960, but in the north the start of significant recent warming is delayed up to 1990.

All the three PCA have almost common PC1 component for the analyzes 1910-2017, i.e., gradual increase of temperature until 1940s, period of declining towards the end of 1950s, a flat phase until the second half of 1970s and steep rise after that. However, the continental PC1 explains only 75.2 \% of the variation of the data, while zonal and seasonal PC1s explain 81.7 \% and 87.6 \% of the corresponding data, respectively.

\end{abstract}

\keywords{Temperature anomaly \and Trend test \and Principal component analysis \and Change-point analysis}

\section{Introduction}
\label{intro}
Global climate is greatly dominated by the increasing temperature since more than 100 years ago \citep{Cook_2016}. The systematic increase of global temperature was broken during the middle of the 20th century. Temperature increased between 1920-1950 (so-called early 20th century warming, ETCW), especially in the northern hemisphere, but cooled after that until the end of 1960s \citep{Davy_2018, Fu_1999, Hegerl_2018, Johannessen_2016, Rybski_2006,  Schlesinger_1994, Wood_2010, Yamanouchi_2011}. 

There has been a debate on to what extent these temperature changes were due to natural variability and/or external forcing \citep{Bengtsson_2004, Egorova_2018, Hegerl_2018, Meehl_2003, Nozawa_2005, Reid_1997, Suo_2013}. Using a one-dimensional climate-ocean model \citeauthor{Reid_1997} came to conclusion that solar and anthropogenic greenhouse-gas forcing made roughly equal contributions to the rise in global temperature during the ETCW. \citeauthor{Meehl_2003} showed that their model required a combination of solar and anthropogenic forcing to produce the ETCW, while the radiative forcing of the increasing greenhouse-gases was dominant for the recent warming effect. \citeauthor{Suo_2013} presented, based on model simulations and observations, that intensified solar radiation and the absence of volcanic activity during the 1920s-1950s can explain much of the early 20th century warming. The anthropogenic forcing could play a role in getting the timing of the peak warming correct. 
Increased heat inflow in the Barents Sea, or anomalous atmospheric circulation patterns in the northern Europe or north Atlantic could also contribute to warming. They concluded that the early 20th century warming was largely externally forced.

\citeauthor{Bengtsson_2004} state that natural variability was a likely cause for the ETCW, with reduced sea ice cover being crucial for the warming. A robust sea ice vs. air temperature relationship was demonstrated by simulations with the atmospheric model forced with observed SST and sea ice concentrations. Also the climate model analysis of \citeauthor{Nozawa_2005} suggest that trends in the external natural factors, e.g., the recovery from volcanic eruptions and the solar irradiance variability, caused more warming in the early 20th century than anthropogenic factors.
Further investigation of the variability of Arctic surface temperature and sea ice cover was performed by analyzing data from a coupled ocean-atmosphere model. The analysis showed that the simulated temperature increase in the Arctic was related to enhanced wind-driven oceanic inflow into the Barents Sea with an associated sea ice retreat.

The review article of \citeauthor{Hegerl_2018} discusses the observed changes during the ETCW and the underlying causes and mechanisms. Attribution studies estimate that about a half of the global warming from 1901 to 1950 was forced by a combination of increasing greenhouse gases and natural forcing, offset to some extent by aerosols. Natural variability also made a large contribution, particularly to regional anomalies like the Arctic warming in the 1920s and 1930s.
\citeauthor{Egorova_2018} made a comprehensive study of ETCW using their atmosphere-ocean chemistry-climate model driven by different combinations of forcing agents. They found a 0.3 $^{o}C$ global warming during 1910-1940, which is about 0.1 $^{o}C$ lower than the observed warming. They, furthermore, found that about half of the explained warming was caused by well-mixed greenhouse-gases, and about one third by the solar UV, visible and infrared irradiance in $250-4000$ nm \citep{Shapiro_2011}.


The global cooling after the ETCW is an interesting phenomenon. Although the greenhouse-gases were still rising, the temperatures decreased, especially in the NH in 1960s and 1970s. \citeauthor{Hodson_2014} suggest that the most likely drivers for the cooling are the "Great Salinity Anomaly" of the late 1960s \citep{Belkin_1998}, the earlier warming of the sub-polar North Atlantic, which may have led to a slowdown in the Atlantic meridional overturning circulation \citep{Robson_2014} and the increase in anthropogenic aerosols, especially sulfur dioxide emissions \citep{Booth_2012, Haywood_2000, Ohmura_2009}.

As to global warming, \citeauthor{Callendar_1938} predicted a 0.03 degrees/decade rise in temperature due to the increasing amount of carbon dioxide \citep{Hawkins_2013}. In the late 1950s it was understood that the concentration of carbon dioxide was indeed increasing in the atmosphere and it was suggested that, sooner or later, this would affect climate \citep{Harris_2010, Revelle_1957}. \citeauthor{Sawyer_1972} predicted global warming and calculated the rate of warming until 2000 \citep{Nicholls_2007}. In the late 1970s it was widely understood that the previous cooling was temporary and that climate warming caused by greenhouse gases, especially $CO_{2}$, would dominate over the cooling effect of smog, aerosols and volcanic dust in the future \citep{Hansen_1981, Robock_1995, Mann_1998, Myhre_2009, Friedman_2013}.

In this paper we study continental, zonal and seasonal land air temperature anomalies in 1880-2017 using principal component (PC) analysis and reverse arrangement (RA) trend analysis. Furthermore, we find greatest change-points in the temperature anomalies combined with their profiles. Particular attention is paid to the role of ETCW in the evolution of temperatures during the 20th century, because it has been largely disregarded in the studies global warming. This paper is organized as follows. Section 2 presents the data and methods used in this study. In Section 3 we analyze continental temperature anomalies and in Section 4 zonal temperature anomalies. In Section 5 we discuss hemispheric seasonal anomalies and give our conclusions in Section 6.

\section{Data and methods}
\label{sec:1}

\subsection{Data}
\label{sec:2}
We use annual and monthly averaged land temperature anomaly data from NOAA (National Oceanic and Atmospheric Administration) web-site for continental data analysis \citep{Menne_2018}, and NASA Goddard Space Flight Center GISTEMP (GISS Surface Temperature Analysis) data for hemispheric, zonal and seasonal land temperature analysis for 1880-2017 \citep{Lenssen_2019}. In zonal analysis both hemispheres are divided into four latitude zones 0-24 , 24-44, 44-64 and 64-90 degrees \citep{Willmott_2018}. The SH zone S90-S64 (Antarctica), however, has been recorded since 1903 only by few stations and is omitted from this analysis.

Annual temperature anomalies (in $^{o}C$) are calculated from the corresponding annual temperature mean values by subtracting the average base period value of the temperature from the temperature values. Similarly monthly (seasonal) anomalies are calculated by subtracting the average value of the corresponding month (season) of the base period from the monthly (seasonal) temperature values. The base periods used in this study are 1910-2000. However, we note that the base period does not affect results provided in this paper, it only changes the level of the zero point of the anomalies. For cumulative sums (later called anomaly profile) the base period is, of course, the whole time series. Furthermore, principal component analysis and reverse arrangement trend test are independent of the base period. For seasons we use conventional definition: northern winter DJF (December, January, February), northern spring MAM (March, April, May), northern summer JJA (June, July, August) and northern fall SON (September, October, November).

\subsection{PCA method}
\label{sec:3}
Principal component analysis has been used in many fields of science, e.g., in chemometrics \citep{Bro_2014}, in data compression \citep{Kumar_2008}, in information extraction \citep{Hannachi_2007}, and in studying the solar and geomagnetic data \citep{Bhattacharyya_2015, Holappa_2014, Takalo_2018}. For a large number of correlated variables PCA finds combinations of a few uncorrelated variables that describe the largest possible fraction of variability in the data. It should be mentioned that many climate indices, e.g., the Atlantic multidecadal oscillation (AMO) and Atlantic oscillation (AO) are defined as modes of PC analysis \citep{Baez_2013, Schlesinger_1994, Trenberth_2006}. (For more information of the method see \citep{Takalo_2018}).


\subsection{Anomaly profile analysis}
\label{sec:4}

For time series of temperature anomalies $X\left(t_{i}\right), i=1...N$, let $\overline{X}$ be the mean value of the anomalies of the whole investigated interval $X\left(t_{i}\right)$. We calculate the cumulative sum time series, i.e., sum of the deviations from the mean anomaly until time $t_{j}$
\begin{equation}
 Y\left(t_{j}\right) = \sum^{t_{j}}_{i=1}\left(X\left(t_{i}\right) - \overline{X}\right).
 \label{eq:cumsum}
\end{equation}

We use here this cumulative sum of the deviation of the anomalies for analysis and call it the "anomaly profile".

\subsection{Reverse arrangement test}
\label{sec:5}

We use the following method for finding trends in time series. Let us have discrete time series $\left\{x_{i}\right\}, i=1, 2..., N$. Beginning from the first value $x_{i}=x_{1}$, we then count the number of times that $x_{i}\!>\!x_{j}$ for $i\,<\,j = 2, 3,...,p$. Each such inequality is called a \textit{reverse arrangement} (RA). This process is then repeated for $x_{i}=x_{2}$, $x_{3}$, ..., $x_{p-1}$. Let the total number of reverse arrangement be $A_{p}$. If the sequence 
$\left\{x_{i}\right\}$ 
is a set independent random variables, then the number of reverse arrangements, i.e. $A_{p}$, is a random normally distributed variable with mean 

\begin{equation}
	\mu_{A_{p}} = \frac{p\left(p-1\right)}{4}
\end{equation}
\begin{flushleft}
and variance
\end{flushleft}
\begin{equation}
	\sigma^{2}_{Ap} = \frac{2p^{3}+3p^{2}-5p}{72}  .
\end{equation}

The $z$-score of $A_{p}$ (also called standardized $A_{p}$) can then be calculated using the following equation \citep{Beck_2006, Bendat_2000, Kendall_1967, Siegel_1988}
\begin{equation}
  z_{p} = \frac{A_{p}\,-\,\mu_{Ap} }{\sigma_{Ap}}.
	\label{eq:z_score}           
\end{equation}

The null hypothesis of the test is that the data points in the time series $\left\{z_{p}\right\}, p=2, 3,..,N$ are independent observations from a random variable. The alternative hypothesis is that the data points are related and show that there is a significant trend underlying the original time series $\left\{x_{i}\right\}, i=1,2...,N$. The 95\% and 99\% acceptance criteria for null hypothesis are that $-1.96 < z < 1.96$ and $-2.58 < z < 2.58$, respectively. A $z$-value outside these intervals means rejection of the null hypothesis with p$<0.05$ and p$<0.01$, respectively.

\subsection{Finding change-points}
\label{sec:6}

We find points, where time series, $\left\{x_{i}\right\}, i=1, 2..., N$, has greatest change in mean values by minimizing the functional \citep{Lavielle_2005, Killick_2012}

\begin{equation}
	J\left(k\right) = \sum^{k-1}_{i=1}\left(x_{i}-\left\langle x\right\rangle^{k-1}_{1}\right)^{2} + \sum^{N}_{i=k}\left(x_{i}-\left\langle x\right\rangle^{N}_{k}\right)^{2} ,
\end{equation}
i.e., finding
\begin{equation}
\min_{k} \left\{\left(k-1\right) var\left([x_{1},x_{2},...,x_{k-1}]\right) + \left(N-k+1\right) var\left([x_{k},x_{k+1},...,x_{N}]\right)\right\}.
\end{equation}

\section{Temperature anomalies of continents}
\label{sec:7}

\begin{figure}
	\centering
	\includegraphics[width=0.8\textwidth]{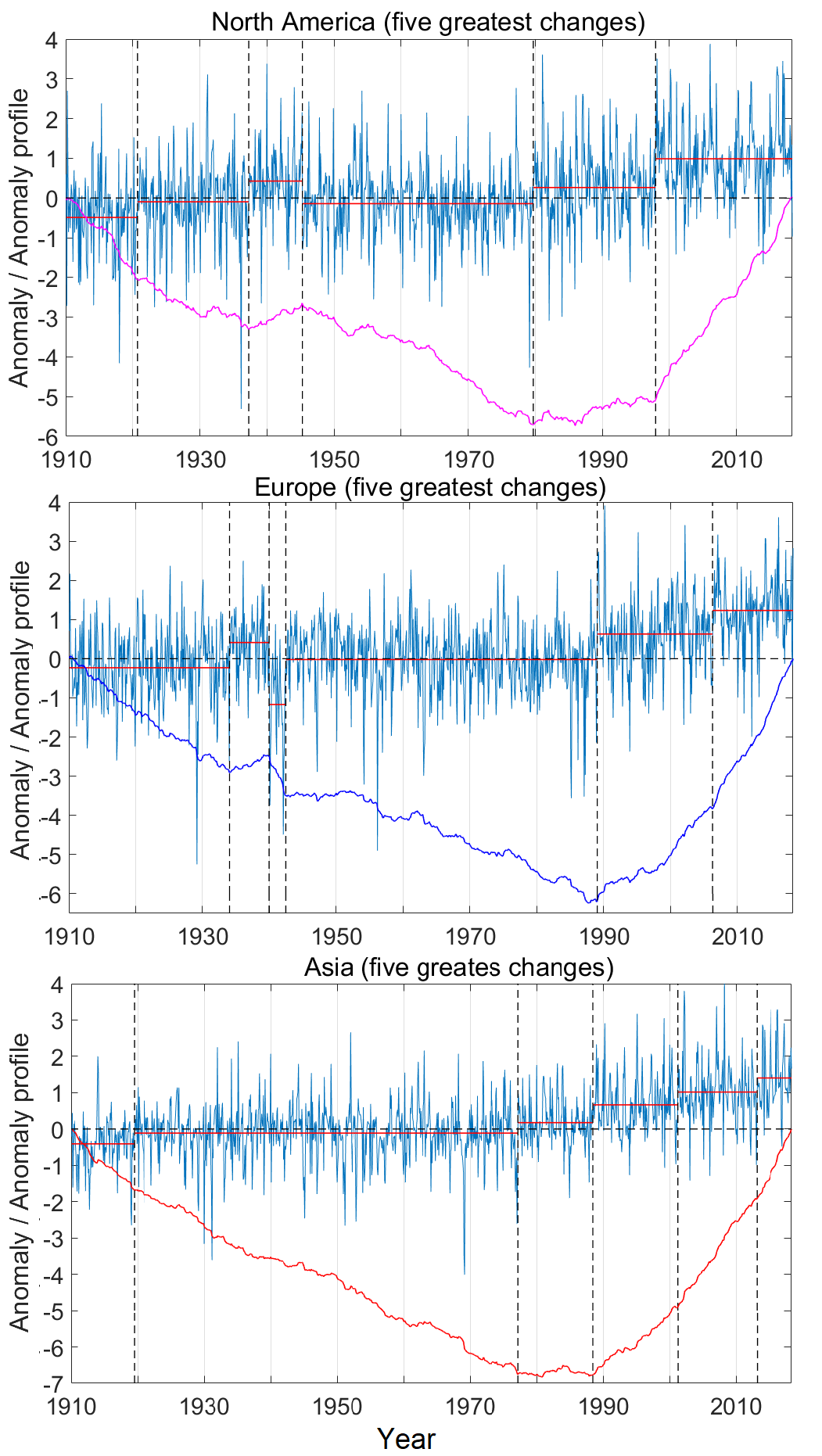}
		\caption{Monthly anomalies and anomaly profiles of the North America, Europe and Asia. The black dashed lines show the greatest changes in anomalies and vertical red lines show the average of the anomalies between change lines. (The anomaly profile in all figures is divided by 40 to fit the same figure with anomalies themselves.)}
		\label{fig:Anomaly_profiles_northern_hemisphere}
\end{figure}

Figure \ref{fig:Anomaly_profiles_northern_hemisphere} shows the monthly anomalies and their profiles for North America, Europe and Asia for 1910-2017. (For continents and zones we divide the profile with 40 to better adjust the anomaly and its profile to same figure). We chose here to present maximum five change-points to better find the differences between continents.The simplest development of the temperature has been in Asia, which have continuous rise after 1910, but been stronger after late 1980s. In North America there was an ETCW between about 1937-1945, after which there was a cooler interval to the late 1970s. The warming period started gradually in 1980s but has been steeper since late 1990s (total variance during 1910-2017 is 1.10 for North America) . The temperature in Europe has been very similar to North America (variance 1.07), but the warming period was earlier, starting about 1934, and ending 1940. After that, simultaneously with second world war, Europe experienced very cold period, which lasted three years. The recent warming in Europe started at the end of 1980s.

\begin{figure}
	\centering
	\includegraphics[width=0.8\textwidth]{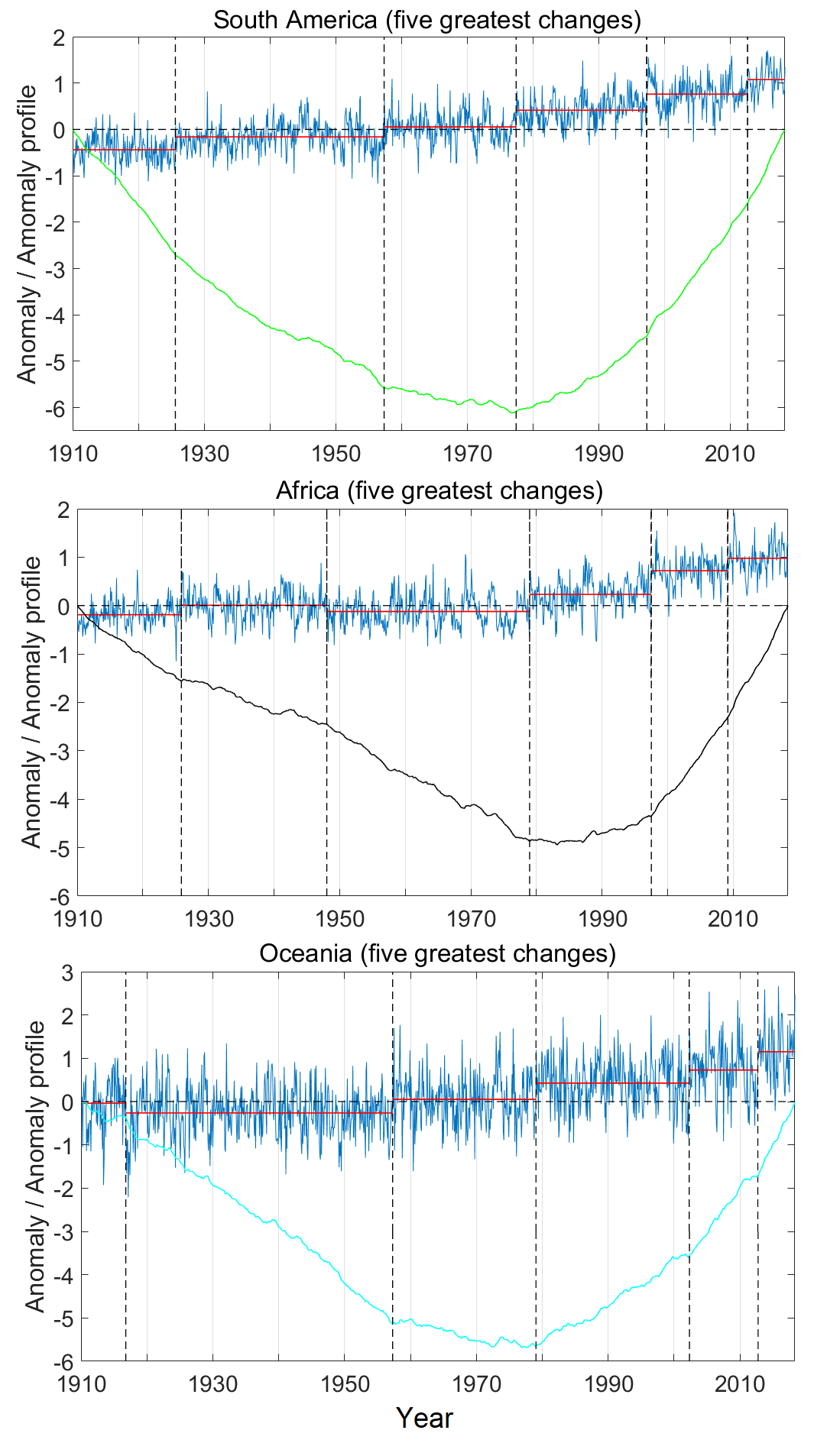}
		\caption{Same as Fig.\ref{fig:Anomaly_profiles_northern_hemisphere}, but for South America, Africa and Oceania.}
		\label{fig:Anomaly_profiles_southern_hemisphere}
\end{figure}

Figure \ref{fig:Anomaly_profiles_southern_hemisphere} shows the monthly anomalies and their profiles for South America, Africa and Oceania (mainly Australia and New Zealand) for 1910-2017. The temperature development of South Africa is very similar to that of Asia. However, the variance of the anomalies is much smaller in South America, i.e, 0.83 for Asia and 0.29 for South America. Also the rise of the temperature is much smoother in South America with only moderate increase recently. Africa has even smaller variance, 0.23, than for South America. Africa had cooler period between 1936-1948, but started warming after that. The recent warming in Africa started in the beginning of 1980s. Oceania, which has largest variation in the anomalies for southern hemisphere continents (variance 0.50) had short warmer period in the 1910s, and after that somewhat cooler until late 1950s. Oceania has had similar moderate increase in temperature after that as South America.

\begin{figure}
	\centering
	\includegraphics[width=1.0\textwidth]{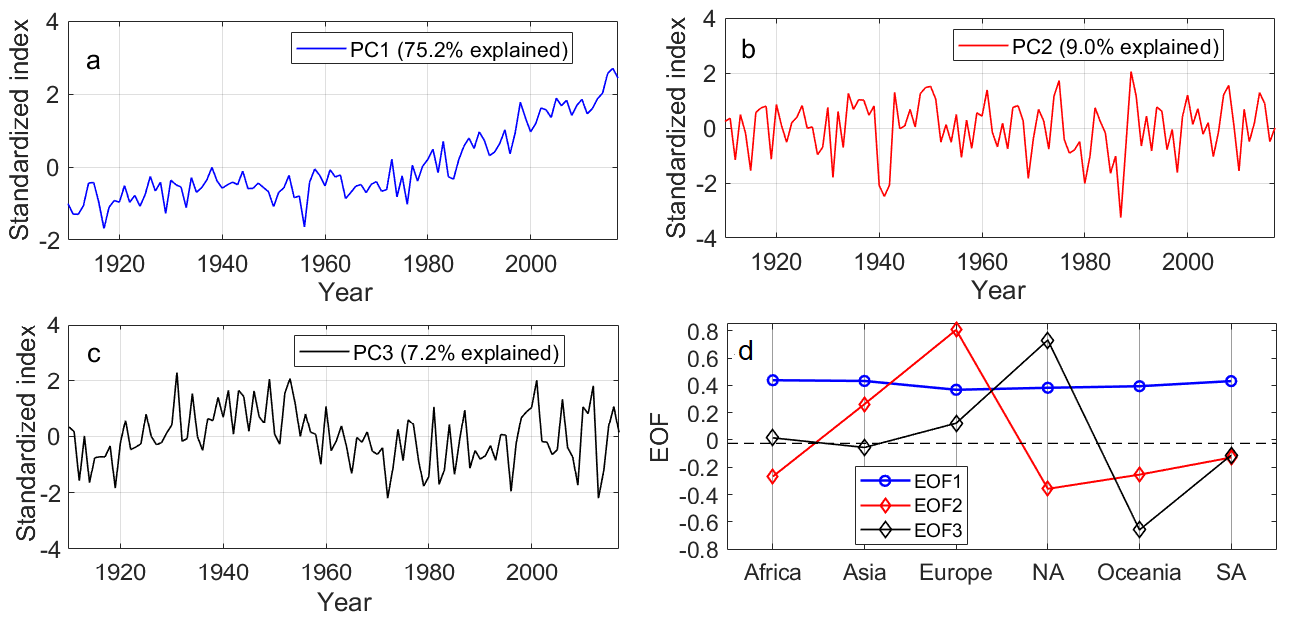}
  \caption{a) PC1 b) PC2 and c) PC3 of the PCA of continental yearly temperature anomalies.d) The EOFs of the PC1, PC2 and PC3.}
		\label{fig:Continents_PCs}
\end{figure}

We have made a PCA for the yearly averaged temperature anomalies of the six continents. Figure \ref{fig:Continents_PCs}a,b and c show the three first PCs, which explain more than 91\% of total variance (PC1 explains 75.2\,\%, PC2 9.0\,\% and PC3 7.2\,\%). PC1 does not exhibit a strong ETCW phenomenon. After the initial increase, there is only a small decrease from the 1940s until the late 1970s (and an increase thereafter). However, the ETCW is also partly included in PC3, which shows a clear increase from 1910s until 1950s, and a subsequent decrease until 1970s. There is also a recent decline of PC3, which makes PC3 to depict a quasi-periodicity of about 60-70 years. It should be noted that PC3 is related to the linearly detrended AMO index \citep{Enfield_2001, Frajka-Williams_2017} with correlation coefficient 0.411 (p$<10^{-5}$). PC2 has a rather flat trend and mainly consists of short period fluctuations that are largest around 1940 and 1990. PC2 is, however, related to the North Atlantic Oscillation index with correlation coefficient 0.478 (p$<10^{-5}$).

Figure \ref{fig:Continents_PCs}d shows the first three EOFs of the continental PCA. While EOF1 depicts very similar values for all continents, EOF2 is largest (and positive) for Europe and second largest (but negative) for North America. It is evident that PC2 mainly describes the short-term differences between these two continents (with smaller differences between other continents). EOF3 is far from zero only for North America and Oceania, with opposite signs. The large positive EOF3 means that North America has an enhanced temperature with respect to all other continents, especially Oceania, between 1930-1960. Accordingly, the ETCW phenomenon is mainly seen in the North America, and least for Oceania. For Oceania the negative EOF3 implies relatively higher temperatures in 1970-1980 with respect to other continents, which leads to an early start of the recent warming in Oceania. The early start was also seen in South America, the only other continent with a significant negative EOF3. The interconnection between the northern (North America) and southern (South America, Oceania) Pacific hemispheres depicted by the oscillating PC3 is called the Interdecadal Pacific Oscillation (IPO) \citep{Salinger_2001}. We find that this interconnection oscillates at the period of about 60-70 years, and may be related to Pacific Decadal Oscillation \citep{Mantua_2002, Newman_2016}, and to Atlantic Multidecadal Oscillation (AMO) \citep{Polyakov_2010, Schlesinger_1994}.

\begin{figure}
	\centering
	\includegraphics[width=0.8\textwidth]{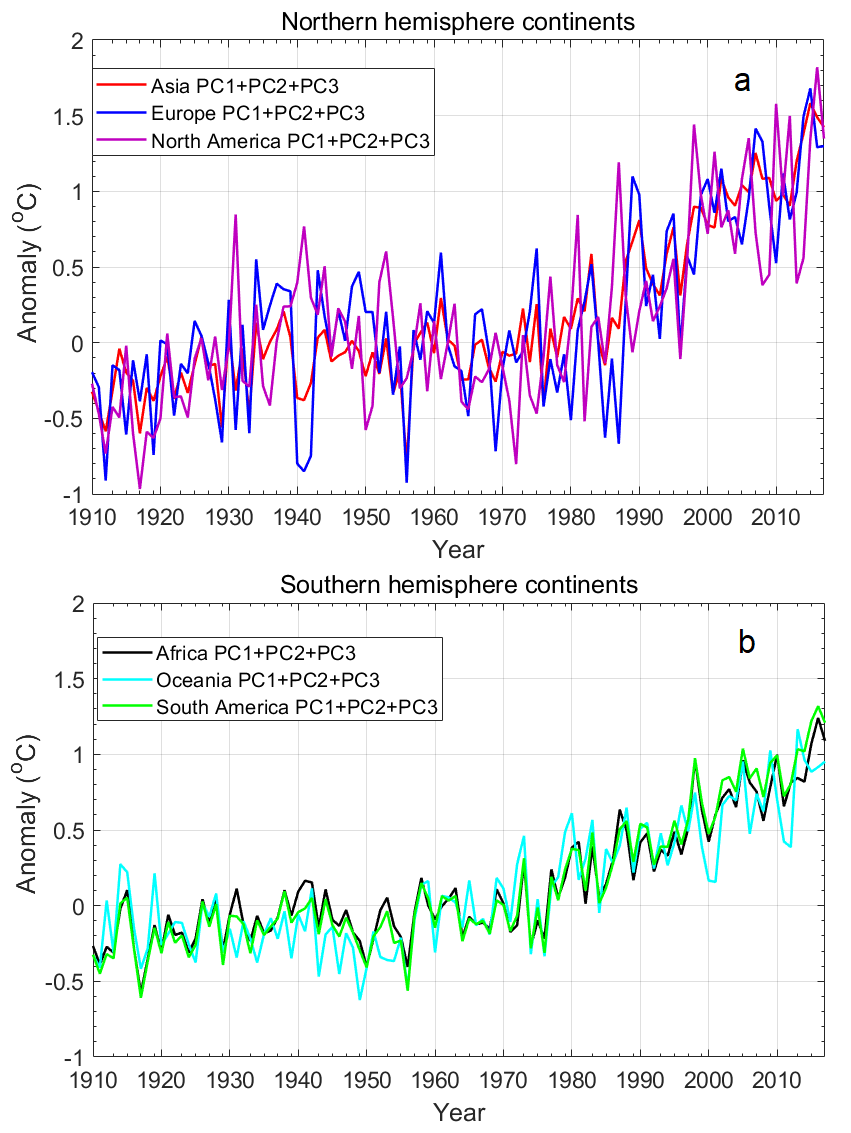}
		\caption{a) PCA proxies (PC1+PC2+PC3) for a) the northern b) southern hemisphere continents in 1910-2017.}
		\label{fig:PC1_PC2_PC3_for_continents}
\end{figure}

Figure \ref{fig:PC1_PC2_PC3_for_continents}a shows the PCA proxies (sums of PC1, PC2 and PC3) for the three northern continents and Figure \ref{fig:PC1_PC2_PC3_for_continents}b for the two southern hemisphere continents and Africa. The PCA proxies of NH continents have larger short-term fluctuations and relative differences than the PCA proxies of SH continents, most likely reflecting the larger fraction of oceans in the southern hemisphere. Note also that many large short-scale fluctuations are in opposite phase between North America and Eurasia (Europe and Asia), especially those around 1940, in 1980 and around 2010. These are among the periods when the corresponding PC2 (see Fig. \ref{fig:Continents_PCs}b) shows the largest values. 

\begin{figure}
	\centering
	\includegraphics[width=0.9\textwidth]{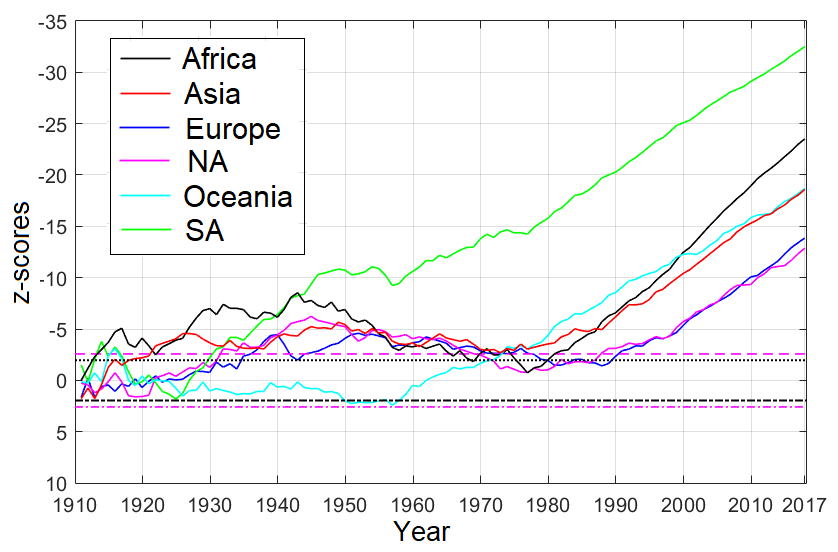}
		\caption{z-scores of the reverse arrangements for continental anomalies. (Note the reversed vertical axis). The 95\,\% (99\,\%) confidence limits are denoted by black (magenta) lines.}
		\label{fig:Trend_tests_of_continents}
\end{figure}

Figure \ref{fig:Trend_tests_of_continents} shows the reverse arrangement tests of the trend for all the six continents. It is obvious that the z-score of South-America diverges from the 99\% confidence level already at about 1930. Moreover, the divergence after 1960 is approximately linear in time with p-value decreasing much below 0.01. This means that in South-America the warming in the 1930s is part of a fairly systematic rise with no clear subsequent temperature decrease, nor ETCW.
Another continent, which differs from others in Figure \ref{fig:Trend_tests_of_continents} is Oceania. Its z-score starts decreasing around 1960 but stays within p$<0.01$ until 1970s. This coincides with the time when the temperature in New Zealand started rising during 1960s \citep{Salinger_1975}. Accordingly, no ETCW is observed in Oceania. The z-scores of the other continents behave quite similarly until about 1980 when Asia and Africa diverge conclusively below p$<0.01$, while Europe and North-America diverge only at about 1990. All these four continents depict a significant warming during the ETCW. The maximum of ETCW is in the 1940s, but the exact timing and length of the ETCW varies between the continents.

\section{Zonal temperature anomalies}
\label{sec:8}

\begin{figure}
	\centering
	\includegraphics[width=1.0\textwidth]{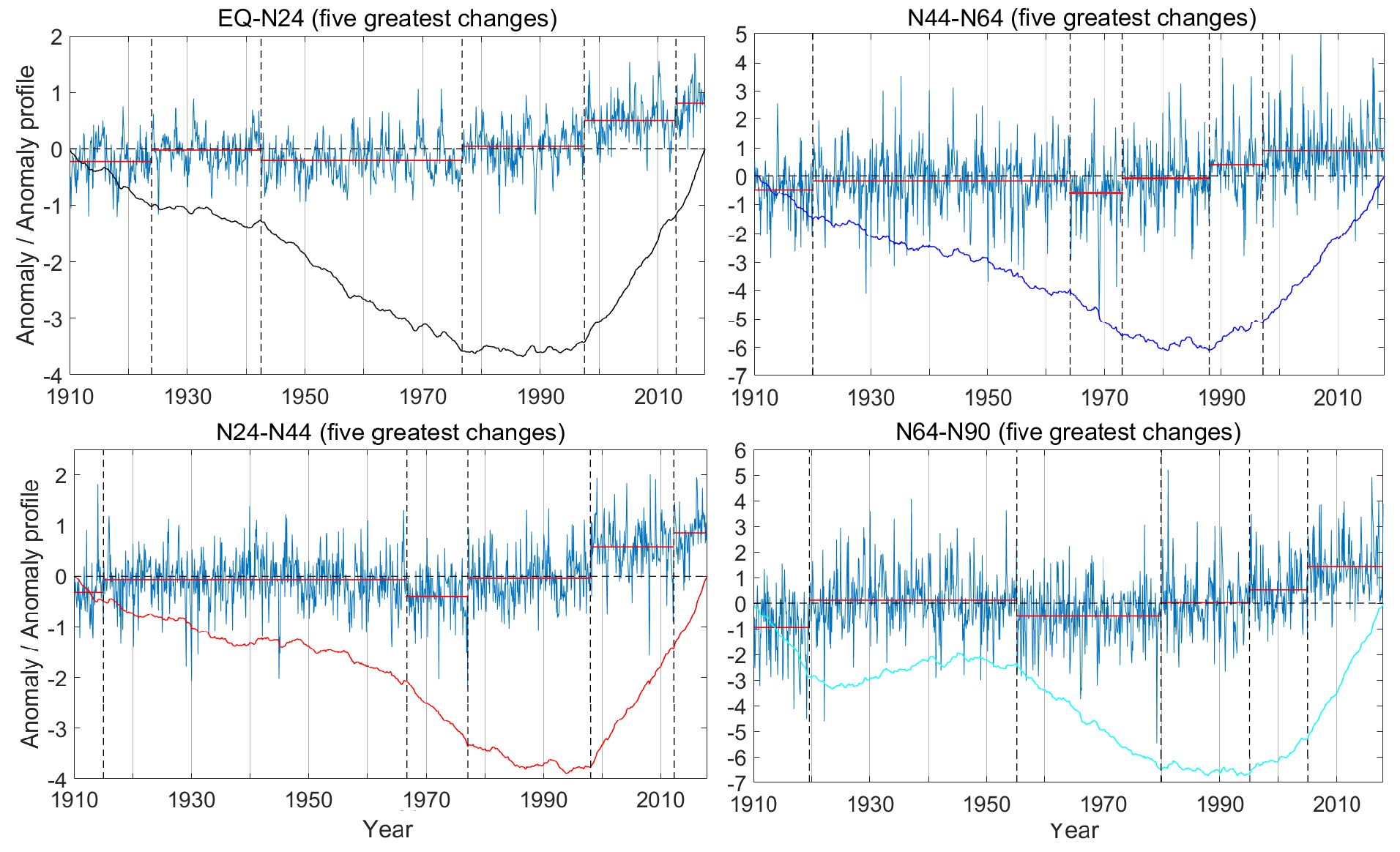}
		\caption{a) NH zonal monthly anomalies and anomaly profiles with five greatest change-points of the anomaly.}
		\label{fig:Zonal_anomaly_profiles_north}
\end{figure}

\begin{figure}
	\centering
	\includegraphics[width=1.0\textwidth]{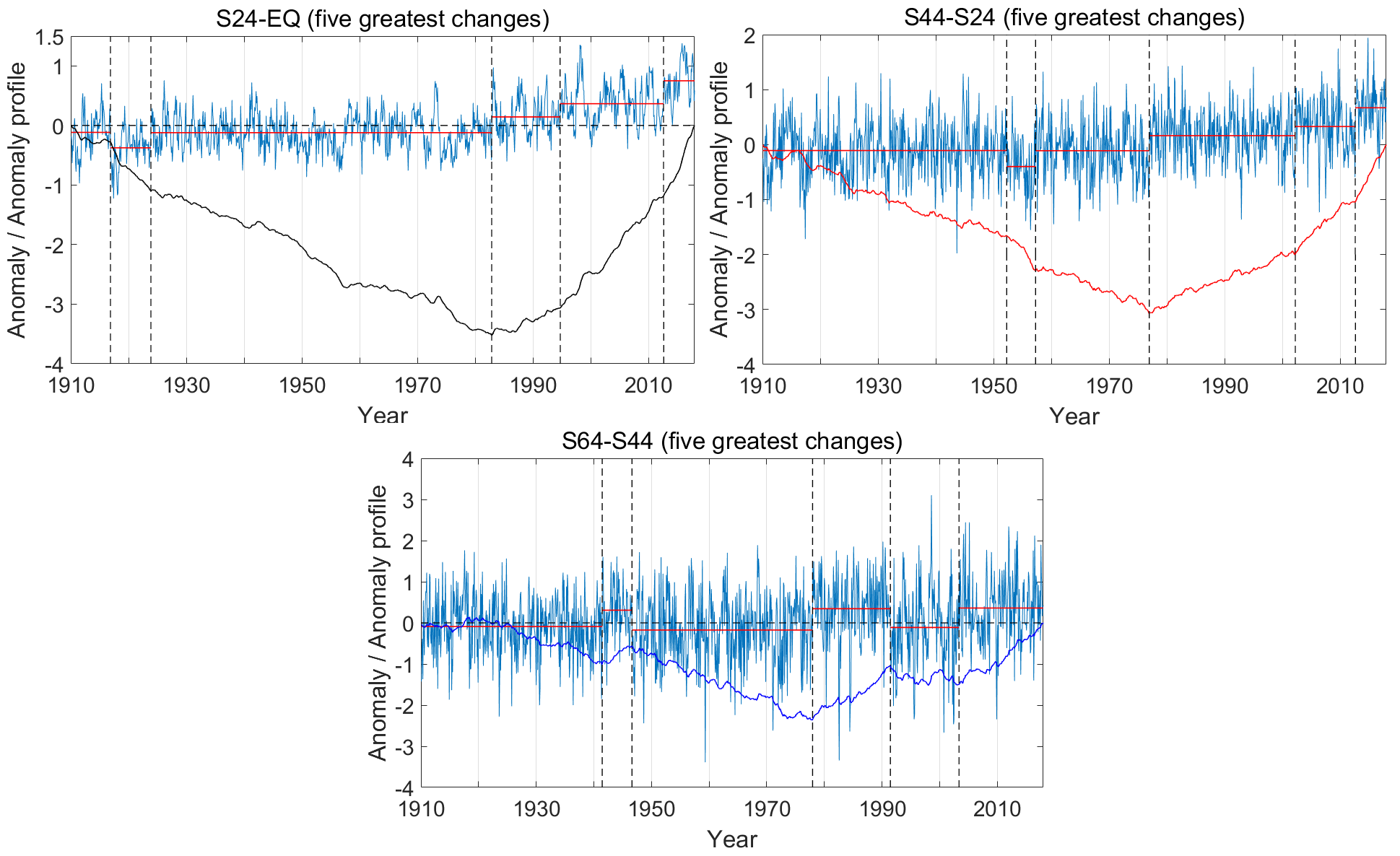}
		\caption{a) SH zonal monthly anomalies and anomaly profiles with five greatest change-points of the anomaly.}
		\label{fig:Zonal_anomaly_profiles_south}
\end{figure}

Figures \ref{fig:Zonal_anomaly_profiles_north} and \ref{fig:Zonal_anomaly_profiles_south} show the monthly land temperature anomalies and their anomaly profiles for four northern hemisphere and three southern hemisphere latitudinal zones 1910-2017, respectively. Since the anomaly profile depends on the selected period, we limit the
data period to 1910-2017 in order to better compare seasonal profiles with continental profiles. We also show the five greatest changes in anomalies during 1910-2017 for each zone. It is evident that the more north (south) we go the larger (smaller, respectively) is the total range of the anomaly profile. The total variances, however, increase when going further north or south from the equator. The variances are 0.194, 0.320, 1.143 and 1.585 for EQ-N24, N24-N44, N44-N64 and N64-N90, respectively, and 0.161, 0.294 and 0.722 for S24-EQ, S44-S24 and S64-S44, respectively. The ETCW is seen only in the northern hemisphere zones and is strongest in the N64-90. This is the only zone, where anomaly is over the average of the whole period, i.e., above zero line between 1920-1955. However, in S64-S44 there was ETCW-like short warming period 1941-46. This zone is also exceptional such that consecutive changes of temperature are alternating up and downs, and there is only a slight recent warming. The anomaly profile minima are earlier in time, in the 1970s to early 1980s, for the southern zones, while the minima for the three northernmost zones are found some 10 years later.

\begin{figure}
	\centering
	\includegraphics[width=1.0\textwidth]{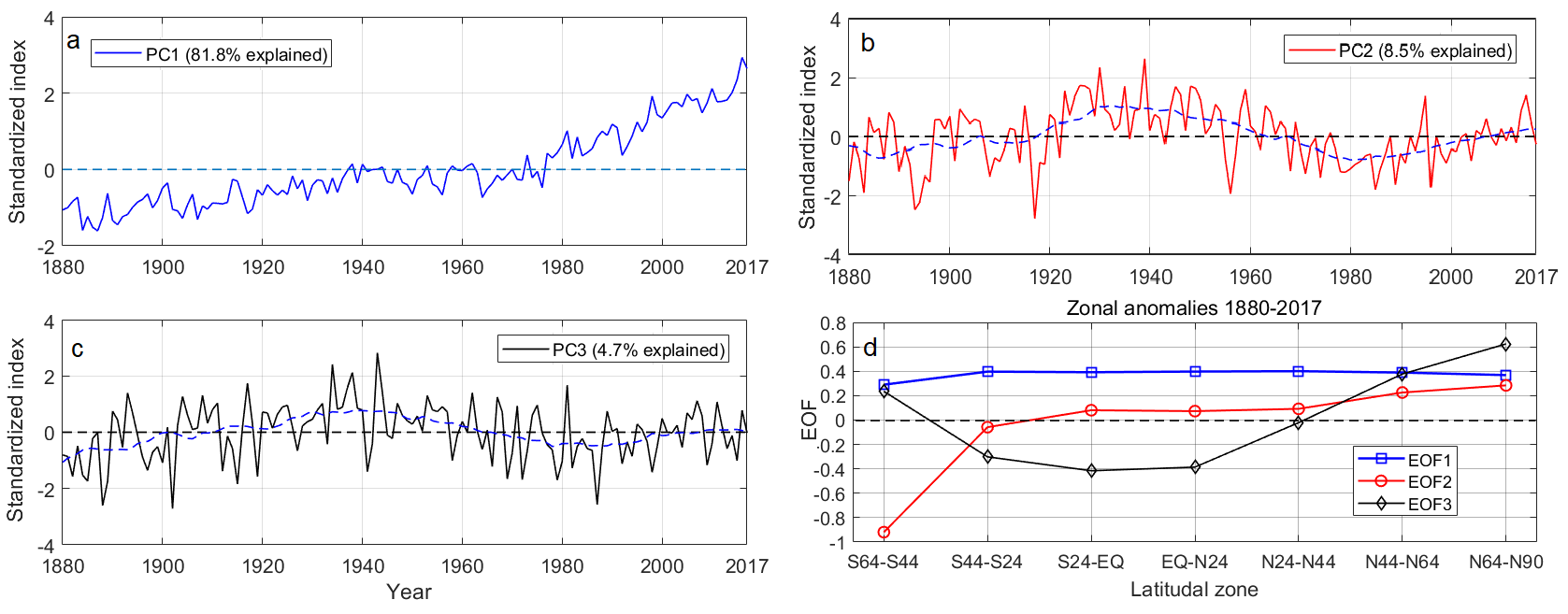}
		\caption{a) PC1, b) PC2 and c) PC3 of zonal temperature anomalies. d) EOFs of zonal PC1, PC2 and PC3.}
		\label{fig:Zonal_anomaly_PCs}
\end{figure}

Figure \ref{fig:Zonal_anomaly_PCs}a, b and c show the first three PCs for the seven zonal yearly averaged anomalies over the whole extent 1880-2017, respectively. PC1 explains now 81.8\,\%, PC2 8.5\,\%, and PC3 4.7\,\% of the total variance. Note, that the ETCW is quite a weak and short maximum in PC1. However, both PC2 and PC3 include a fairly strong ETCW evolution (we show also 21 year trapezoidal smoothed curves of PC2 and PC3). Figure \ref{fig:Zonal_anomaly_PCs}d shows the related EOFs. EOF1 has an almost equal power in all zonal anomalies, only somewhat smaller for S64-S44. Zone S64-S44 has a very large negative value of EOF2, while zone N64-N90 has the largest positive value. It is clear that PC2 aims to characterize the difference between S64-S44 from the other zones, in particular from the two northernmost zones. Similarly, PC3 takes into account the difference between the most poleward zones (on either hemisphere) and the equatorial (and low southern) zones. Note also that, because of the similar long-term variation of PC2 and PC3, the ETCW evolution is largely suppressed for those zones whose EOF2 and EOF3 have opposite signs. This is true for the southernmost zone and the two equatorial zones.
As to the above discussion, it is no wonder that the correlation coefficients between zonal PC2 and AMO index is 0.380 (p=0.000049) and PC3 and SOI index is 0.479 (p$<10^{-5}$). Note, especially, that EOF3 is large negative for S44-S24, S24-EQ and EQ-N24.

\begin{figure}
	\centering
	\includegraphics[width=0.9\textwidth]{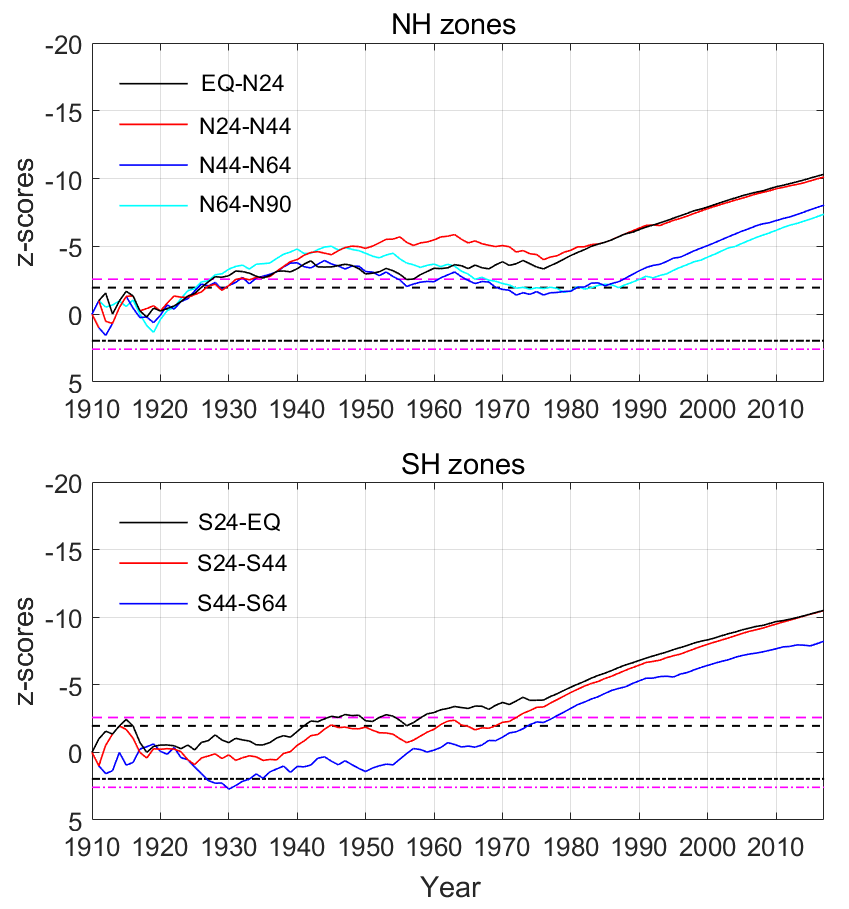}
		\caption{z-scores of the reverse arrangement test for a) the northern b) the southern hemisphere zonal temperatures. (Note reverse vertical axes.) The 95\,\% (99\,\%) confidence limits are denoted by black (magenta) lines.}
		\label{fig:Zonal_trends}
\end{figure}

Figures \ref{fig:Zonal_trends}a and \ref{fig:Zonal_trends}b show the trend analysis for northern and southern hemisphere zones, respectively. We restricted the trend analysis again to the period 1910-2017 so that it is comparable to the earlier trend analyzes. Note that, as for the seasonal analysis, the ETCW is significant in all NH zones. In the two lower-latitude NH zones the ETCW warming continues as a significant increase until the recent warming. However, higher in the north the cooling after the ETCW is larger and the recent warming is delayed to start until about 1990. No significant warming is found during the ETCW interval in the SH zones, although marginal warming is seen in S24-EQ in the 1940s. Recent warming becomes significant in S24-EQ around 1960, in the two other SH zones in the 1970s.

\section{NH and SH seasonal anomalies}
\label{sec:9}

\begin{figure}
	\centering
	\includegraphics[width=1.05\textwidth]{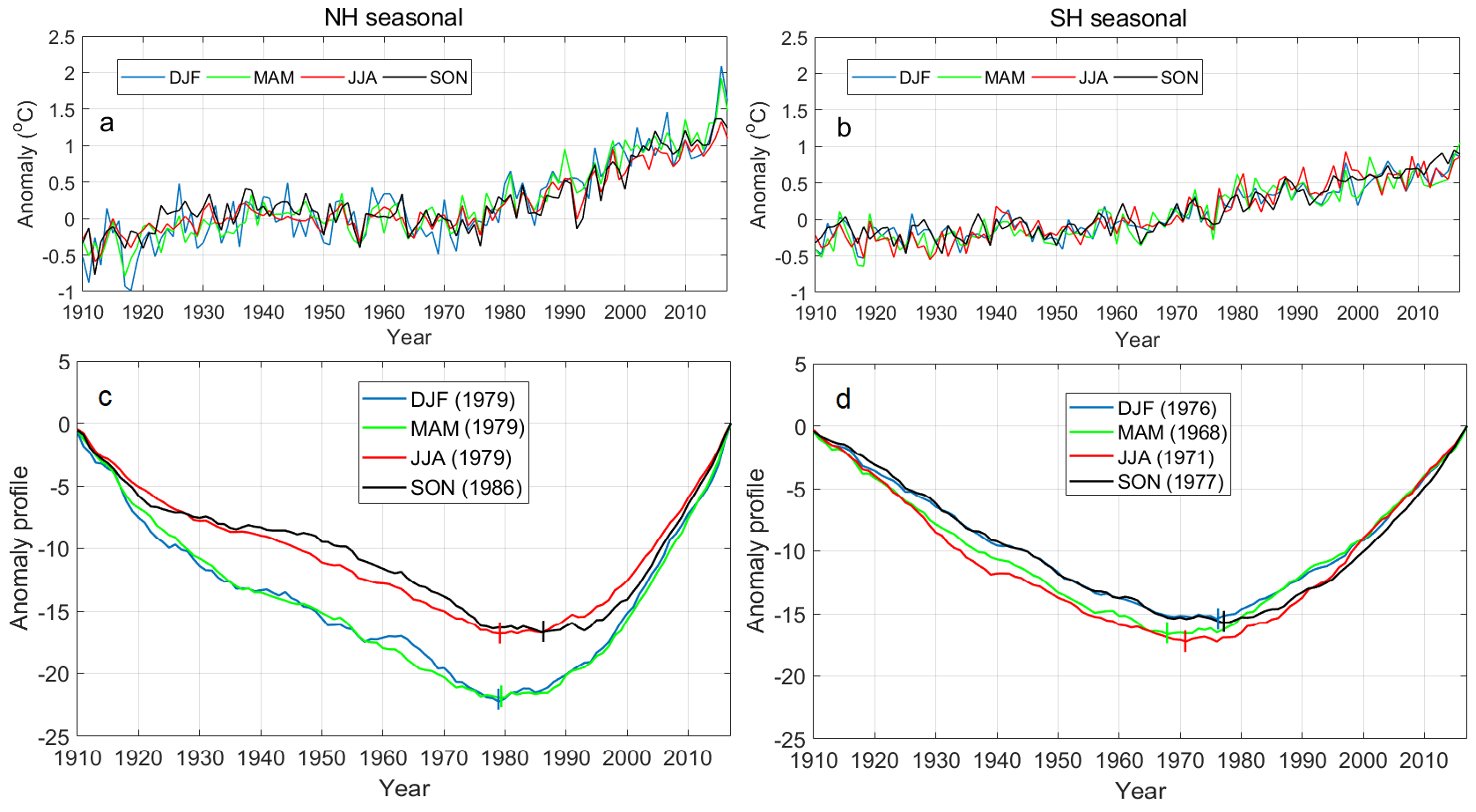}
		\caption{a) Northern hemisphere seasonal anomalies and c) their profiles. b) Southern hemisphere seasonal anomalies and d) their profiles. Small vertical bars in c) and d) show the minimum years of the continental cumulative anomalies, also given in parenthesis in the legends.}
		\label{fig:Seasonal_anomaly_profiles}
\end{figure}

We have made a PC analysis for NH and SH seasonal land temperature anomalies. Figure \ref{fig:Seasonal_anomaly_profiles} shows the temperature anomalies and anomaly profiles for each season, NH on panels a and c, and SH on panels b and d, respectively. Since the anomaly profile depends on the selected period, we limit the data period to 1910-2017 in order to better compare seasonal profiles with continental profiles. (We do not try to find change-points here, because we have, of course, only yearly data for seasons, and consequently the resolution is too sparse to make reliable analysis). Figure \ref{fig:Seasonal_anomaly_profiles} shows that most seasonal anomalies of the southern hemisphere reach their minima at about 1970, about ten years earlier than most of the northern hemisphere seasons. As for the above presented PCA of continents, this is mainly due to the ETCW and subsequent cooling period in the NH, which delays the start of the recent warming in the northern hemisphere. Note also that the NH seasons differ considerably from each other with ETCW being more strongly present in the summer and fall seasons of the NH.
 
\begin{figure}
	\centering
	\includegraphics[width=0.8\textwidth]{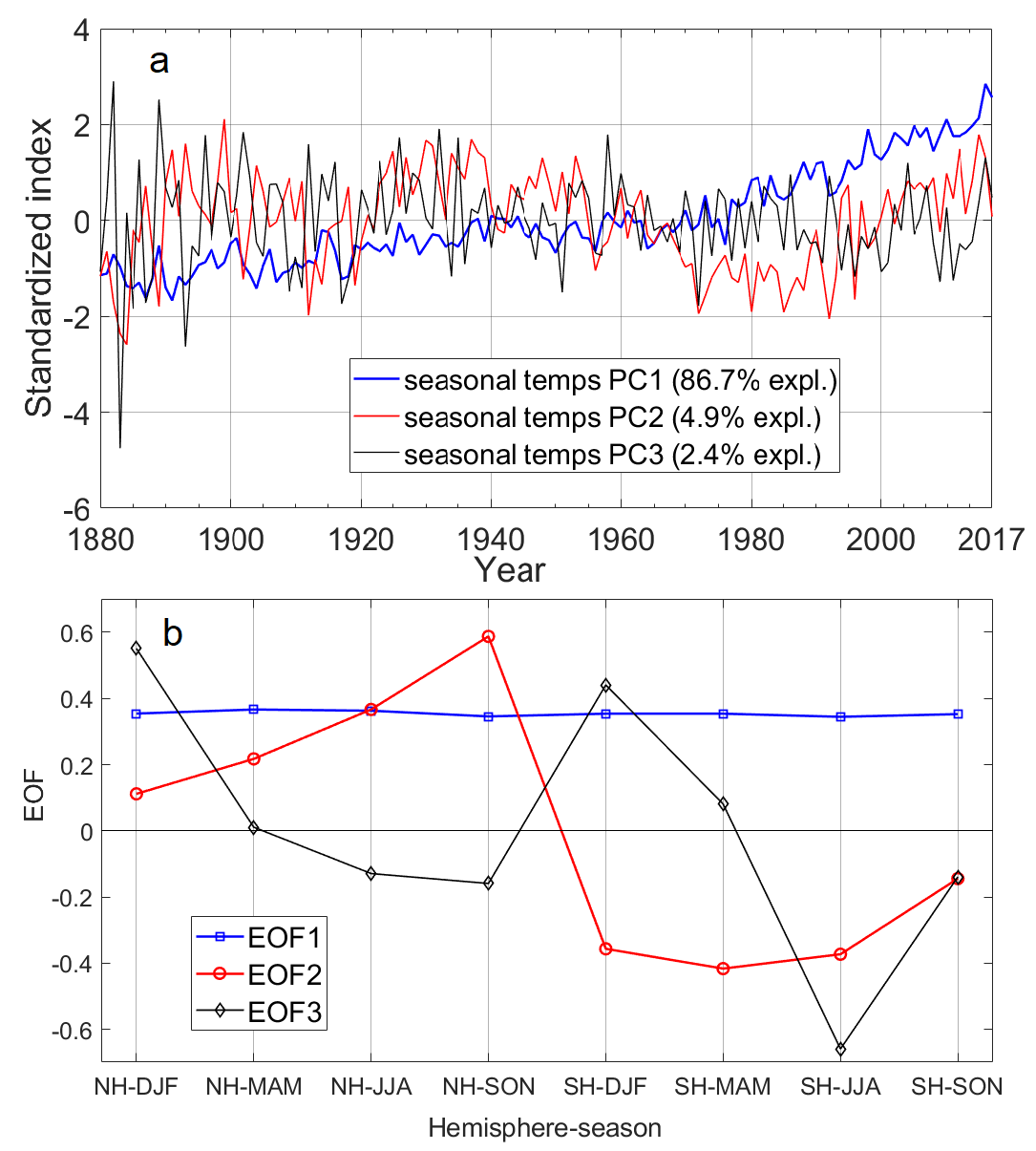}
		\caption{a) PC1, PC2 and PC3 of the hemispheric seasonal anomaly PCA. b) Corresponding EOF1, EOF2 and EOF3.}
		\label{fig:Seasonal_PCs} 
\end{figure}

Figure \ref{fig:Seasonal_PCs}a shows the three leading PCs for the seasonal temperature anomalies using the full length of seasonal data in 1880-2017. PC1 accounts for as much as 86.6\%, PC2 4.9\% and PC3 2.4\% of the total variance. The high percentage of PC1 means that all eight seasonal anomalies follow quite a similar temporal evolution of PC1, where temperature rises systematically from 1880 to a weak maximum of the ETCW around 1940, with only a short subsequent decrease lasting until the late 1950s. 
PC2 again describes the different fraction of ETCW in the different seasons. Figure \ref{fig:Seasonal_PCs}b shows the corresponding EOFs with NH and SH EOF2s having opposite signs. The highest positive EOF2 is found in NH fall, while the other three seasons in the SH have the largest negative EOF2s. That is why it is not surprising that PC2 is correlated with linearly detrended AMO index with coefficient 0.557 ( p$<10^{-8}$).

Seasonal anomaly PC3 accounts only for 2.4\% of the variance of seasonal anomalies. Note that the corresponding EOF3 is highly positive during both north and south DJF. Although PC3 looks quite noisy, it correlates significantly with wintertime (DJF) NINO34 with coefficient 0.358 (p=0.00015). The correlation with spring (MAM) NINO34 component is still 0.246 (p=0.011), but for other seasons insignificant. This reminds us about the origin of naming El Ni{\~n}o phenomenon, which was first noticed in winter and called "El Ni{\~n}o de Navidad" (Christmas Child) by the fishermen \citep{Winchester_2017}.


\begin{figure}
	\centering
	\includegraphics[width=0.9\textwidth]{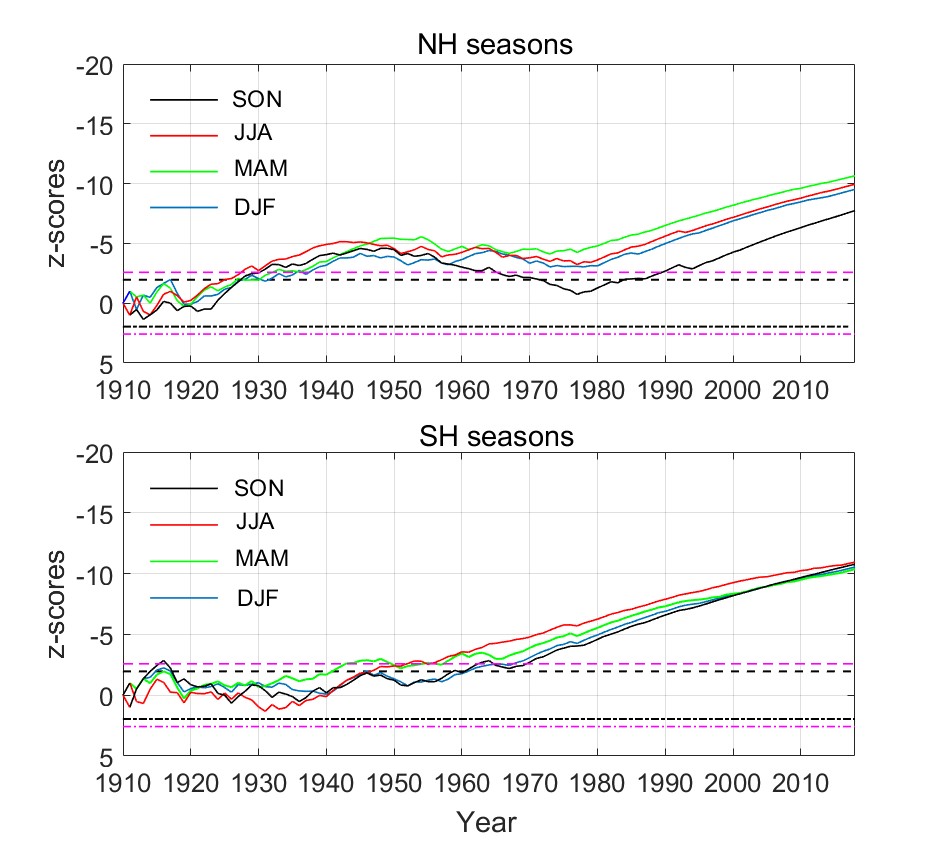}
		\caption{z-scores of the reverse arrangements for a) the northern b) the southern hemisphere seasons. (Note reverse vertical axes). The 95\,\% (99\,\%) confidence limits are denoted by black (magenta) lines.}
		\label{fig:Seasonal_trends} 
\end{figure}

Figures \ref{fig:Seasonal_trends}a and \ref{fig:Seasonal_trends}b show the seasonal z-scores of the reverse arrangements test for northern and southern hemisphere, respectively. Note the significance of the ETCW for all seasons in the northern hemisphere. The NH z-scores move out of the p=0.01 limit in the 1920s to 1930s. In most NH seasons the ETCW continues as a significant rise until the recent warming. However, in NH fall the temperature decreases relatively more after the ETCW, and the recent warming starts only in the 1990s (as a significant trend). The z-scores in the SH decrease fairly systemically since 1930s but remain random (within p=0.01) until the late 1950s to early 1970s depending on season. A short significant ETCW is seen only in spring.

\section{Discussion and conclusions}
\label{sec:10}

We have used the principal component analysis, reverse arrangement (RA) trend test and anomaly profiles to study land temperature anomalies from 1880 onwards. The RA trend test is, as far as we know, used the for the first time in temperature analysis. Also combining anomaly profile (cumsum), with change-points analysis clarifies the temperature evolution remarkably.  

The principal component analysis of the yearly temperature anomalies of the six continents reveals that the continents depict the ETCW very differently. PC1, which explains 75.2\,\% of inter-annual variability includes warming from 1910s to to 1940s, but a very modest  cooling thereafter and a notable recent warming since the 1970s. PC2, which includes 9.0\,\% of the variance between the continents, and is related to NAO index, includes no trend and describes mainly the short-term fluctuations which are in opposite phase between North America and Eurasia (Europe and Asia), especially around 1940 and around 1990. The different appearance of the ETCW in the six continents is mainly included in PC3, which describes 7.2\,\% of variance, and is highly correlated with detrended AMO index.

Trend analysis shows that there is a significant ETCW warming in all other continents except for Oceania.
However, there are differences in the start time and duration of significant warming and the ETCW maximum between these five continents. Most of them have their ETCW maximum in the 1940s.
The EOF3 of Oceania is strongly negative, and its evolution in the early 1900s was almost opposite to ETCW with a decrease of temperature from the early century to a minimum in the 1940s and 1950s. 
In four ETCW continents (Africa, Asia, Europe and North America) the cooling after the ETCW was strong enough to considerably delay the start of recent warming. 
Recent warming started in North America in the late 1980s and in Europe only around 1990. 
In South America there is little cooling before the ETCW, and surprisingly warming is systematically significant since the start of the ETCW around 1930. As far as we know, this slow but continuous rise in the temperature of South America has not been reported earlier. 
We also find that PC3, and thereby the ETCW phenomenon, depict a 60-70-year oscillation, which is related, at least, to AMO. If the AMO index is decreasing after present maximum, PC3 can be used to forecast that there will be a slight cooling in the Northern hemisphere continents, especially North America. Note, however, that PC3 accounts only about 7\% of the whole variation of the anomalies.

The principal component analysis of the temperature anomalies of seven latitude zonal regions (we leave out Antarctica) shows that, as for seasonal data, the PC1 (about 81.8\,\% of variance) depicts fairly systematic warming throughout the studied time interval, with a weak ETCW maximum around 1940 and only a weak cooling thereafter until 1970s. 
However, both PC2 (about 8.7\% of variance) and PC3 (about 4.7\% of variance) include a fairly similar, roughly in-phase evolution. PC2 and PC3 mainly deviate in their different short-term variation. 
According the the EOF2 and EOF3, the ETCW warming is strongest in the two northernmost zones (N44-N64 and N64-N90). 

Trend analysis shows that most NH zones have the ETCW maximum in 1940s, but have mutual differences in the start time and duration of significant warming and the level of cooling after the ETCW.
The southern zones depict no significant ETCW warming, only marginally significant warming in the equatormost zone (S24-EQ), and a short warming period in S64-S44 during the first half on 1940s.
In fact, the southernmost zone (S64-S44) seems to have a cooling phase during the ETCW period, but the trend analysis shows that this is only marginally significant (at p$<0.01$), and strongest around 1930.
The start of significant recent warming varies considerably. 
In the two lowest northern zones (EQ-N24, N24-N44) the warming is significant since the ETCW, and increased warming starts in 1970s. 
In the two northernmost zones (N44-N64, N64-N90) the cooling after the ETCW delays the start of recent warming until around 1990.
In the southern hemisphere, the recent warming starts first closest to the equator in the 1950s and latest in the southernmost zone in the late 1970s. 

We also studied the hemispheric temperature anomalies during four seasons (winter, spring, summer, and fall).
The high percentage of PC1 (about 86.6\,\% of variance) for hemispheric seasons means that all eight seasonal anomalies follow quite a similar temporal evolution. However, the eight seasons are divided by PC2 (about 4.9\,\% of variance) into the four northern hemisphere seasons with positive EOF2 and for southern hemisphere seasons with negative EOF2. PC2 is highly correlated with detrended AMO index, at least in the period 1910-2017, and includes not only the ETCW phenomenon and the subsequent cooling, which ends in the 1970s, but also an earlier warming at the end of 19th century and a cooling in the beginning of 20th century.The PC2 type evolution is strongest in NH fall season and less strong in all other NH seasons.

In trend analysis all NH seasons also depict a significant ETCW warming. While most NH seasons have the ETCW maximum in the 1940s, there are mutual differences in the timing of ETCW and level of subsequent cooling. 
No SH season depict significant warming during the ETCW period. In 1950s a systematic increase in all SH seasons starts, which becomes significant in the late 1950s for spring and summer and in the late 1960s for fall and winter. 
In the NH the start of final warming is delayed until 1990s for NH fall, and 1970s for other seasons.

\begin{figure}
	\centering
	\includegraphics[width=0.8\textwidth]{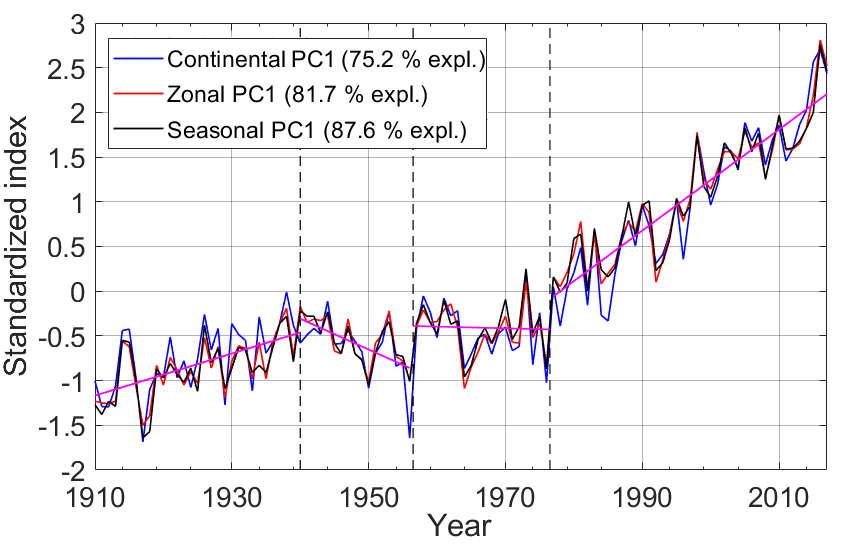}
		\caption{The PC1s of continental, zonal and seasonal anomaly PC analyzes. Magenta lines show the average linear trends and vertical dashed lines the change-points.}
		\label{fig:PC1s} 
\end{figure}

Although we have studied the global temperature evolution during 20th century in three different PCA analysis, there is almost common PC1 for all analyzes (see Fig. \ref{fig:PC1s}). It is evident, that the overall development of the anomalies are very similar. (Here we use PC analyzes for zonal 1910-2017 and seasonal 1910-2017 data). All the PC1s show slight warming towards the end of 1940s, a period of decline in anomalies until second half of 1950s, a rather flat interval until second half of 1970s, and a steep rise after that. There are some minima, which are deeper in the continental PC1. These are, e.g., years 1956-57 and 1984-85, when there were extremely cold winters, at least, in Europe \citep{Twardosz_2016, Dizerens_2017}. These events are smoothed out in the zonal and seasonal PC analyzes. Notice, however, that the continental PC1 explains only 75.2 \% of the variation of the data, while zonal and seasonal PC1s explain 81.7 \% and 87.6 \% of the corresponding data, respectively.

\begin{acknowledgements}
We acknowledge the financial support by the Academy of Finland to the ReSoLVE Center of Excellence (project no. 307411). The annual and monthly averaged continental temperature anomaly data were retrieved from NOAA (National Oceanic and Atmospheric Administration) web-site (https://www.ncdc.noaa.gov/cag/), and data for hemispheric, zonal and seasonal land temperature analysis were retrieved from NASA Goddard Space Flight Center GISTEMP (GISS Surface Temperature Analysis) web-page (https://data.\newline giss.nasa.gov/gistemp/).
\end{acknowledgements}

%
%

\bibliographystyle{spr-mp-sola}      
\bibliography{references_JT}

\begin{thebibliography}{60}
\ifx\bisbn     \undefined \def\bisbn  #1{ISBN #1}\fi
\ifx\binits    \undefined \def\binits#1{#1}\fi
\ifx\bauthor   \undefined \def\bauthor#1{#1}\fi
\ifx\batitle   \undefined \def\batitle#1{#1}\fi
\ifx\bjtitle   \undefined \def\bjtitle#1{\textit{#1}}\fi
\ifx\bvolume   \undefined \def\bvolume#1{\textbf{#1}}\fi
\ifx\byear     \undefined \def\byear#1{#1}\fi
\ifx\bissue    \undefined \def\bissue#1{#1}\fi
\ifx\bfpage    \undefined \def\bfpage#1{#1}\fi
\ifx\blpage    \undefined \def\blpage #1{#1}\fi
\ifx\burl      \undefined \def\burl#1{\textsf{#1}}\fi
\ifx\href      \undefined \def\href#1#2{\textsf{#2}}\fi
\ifx\betal     \undefined \def\betal{\textit{et al.}}\fi
\ifx\bctitle   \undefined \def\bctitle#1{#1}\fi
\ifx\beditor   \undefined \def\beditor#1{#1}\fi
\ifx\bbtitle   \undefined \def\bbtitle#1{\textit{#1}}\fi
\ifx\bedition  \undefined \def\bedition#1{#1}\fi
\ifx\bseriesno \undefined \def\bseriesno#1{\textbf{#1}}\fi
\ifx\blocation \undefined \def\blocation#1{#1}\fi
\ifx\bsertitle \undefined \def\bsertitle#1{\textit{#1}}\fi
\ifx\bsnm      \undefined \def\bsnm#1{#1}\fi
\ifx\bsuffix   \undefined \def\bsuffix#1{#1}\fi
\ifx\bparticle \undefined \def\bparticle#1{#1}\fi
\ifx\barticle  \undefined \def\barticle#1{}\fi
\ifx\binstitute  \undefined \def\binstitute#1{#1}\fi
\ifx\bpublisher  \undefined \def\bpublisher#1{#1}\fi
\ifx\doiurl    \undefined \def\doiurl#1{\href{#1}{\textsf{DOI}}}\fi
\makeatletter
\def\safeHref#1#2#3{\in@{http}{#2}\ifin@\href{#2}{#3}\else\href{#1#2}{#3}\fi}
\makeatother
\ifx\adsurl    \undefined
  \def\adsurl#1{\safeHref{https://ui.adsabs.harvard.edu/abs/}{#1}{\textsf{ADS}}}\fi
\ifx\arxivurl  \undefined
  \def\arxivurl#1{\safeHref{http://arxiv.org/abs/}{#1}{\textsf{arXiv}}}\fi
\ifx\botherref \undefined \def\botherref#1{}\fi
\ifx\url       \undefined \def\url#1{\textsf{#1}}\fi
\ifx\bchapter  \undefined \def\bchapter#1{}\fi
\ifx\bbook     \undefined \def\bbook#1{}\fi
\ifx\bcomment  \undefined \def\bcomment#1{#1}\fi
\ifx\oauthor   \undefined \def\oauthor#1{#1}\fi
\ifx\citeauthoryear \undefined\def \citeauthoryear#1{#1}\fi
\def\endbibitem {}
\ifx\bconflocation  \undefined \def\bconflocation#1{#1} \fi

\bibitem[\protect\citeauthoryear{{Baez} \textit{et~al.}}{2013}]{Baez_2013}
\begin{barticle}
\bauthor{\bsnm{{Baez}}, \binits{J.C.}},
\bauthor{\bsnm{{Gimeno}}, \binits{L.}},
\bauthor{\bsnm{{Gomes-Gesteira}}, \binits{M.}},
\bauthor{\bsnm{{Ferri-Yanez}}, \binits{F.}},
\bauthor{\bsnm{{Real}}, \binits{R.}}:
\byear{2013},
\batitle{{Combined Effects of the North Atlantic Oscillation and the Arctic
  Oscillation on Sea Surface Temperature in the Alborán Sea}}.
\bjtitle{PloS One}
\bvolume{8}(\bissue{4}).
\doiurl{https://doi.org/10.1371/journal.pone.0062201}.
\end{barticle}
\endbibitem

\bibitem[\protect\citeauthoryear{{Beck} \textit{et~al.}}{2006}]{Beck_2006}
\begin{botherref}
\oauthor{\bsnm{{Beck}}, \binits{T.W.}},
\oauthor{\bsnm{{Housh}}, \binits{T.J.}},
\oauthor{\bsnm{{Weir}}, \binits{J.P.}},
\oauthor{\bsnm{{Cramer}}, \binits{J.T.}},
\oauthor{\bsnm{{Vardaxis}}, \binits{V.}},
\oauthor{\bsnm{{Johnson}}, \binits{G.O.}},
\oauthor{\bsnm{{Coburn}}, \binits{J.W.}},
\oauthor{\bsnm{{Malek}}, \binits{M.H.}},
\oauthor{\bsnm{{Mielke}}, \binits{M.}}:
2006,
{An examination of the runs test, reverse arrangements test, and modified
  reverse arrangements test for assessing surface EMG signal stationarity}.
\textit{J. Neurosci. Methods},
242.
\end{botherref}
\endbibitem

\bibitem[\protect\citeauthoryear{{Belkin} \textit{et~al.}}{1998}]{Belkin_1998}
\begin{barticle}
\bauthor{\bsnm{{Belkin}}, \binits{I.M.}},
\bauthor{\bsnm{{Levitus}}, \binits{S.}},
\bauthor{\bsnm{{Antonov}}, \binits{J.}},
\bauthor{\bsnm{{Malmberg}}, \binits{S.A.}}:
\byear{1998},
\batitle{{Great Salinity Anomalies in the North Atlantic}}.
\bjtitle{Prog. Oceanogr.}
\bvolume{41},
\bfpage{1}.
\doiurl{https://doi.org/10.1016/S0079-6611(98)00015-9}.
\adsurl{1998PrOce..41....1B}.
\end{barticle}
\endbibitem

\bibitem[\protect\citeauthoryear{{Bendat} and {Piersol}}{2000}]{Bendat_2000}
\begin{bbook}
\bauthor{\bsnm{{Bendat}}, \binits{J.S.}},
\bauthor{\bsnm{{Piersol}}, \binits{A.G.}}:
\byear{2000},
\bbtitle{{Random Data Analysis and Measurement Procedures}},
\bpublisher{Wiley, New York}, \blocation{???},
\bfpage{96}.
\end{bbook}
\endbibitem

\bibitem[\protect\citeauthoryear{{Bengtsson}, {Semenov}, and
  {Johannessen}}{2004}]{Bengtsson_2004}
\begin{barticle}
\bauthor{\bsnm{{Bengtsson}}, \binits{L.}},
\bauthor{\bsnm{{Semenov}}, \binits{V.A.}},
\bauthor{\bsnm{{Johannessen}}, \binits{O.M.}}:
\byear{2004},
\batitle{{The Early Twentieth-Century Warming in the Arctic-A Possible
  Mechanism.}}
\bjtitle{J. Clim.}
\bvolume{17},
\bfpage{4045}.
\doiurl{https://doi.org/10.1175/1520-0442(2004)017$<$4045:TETWIT$>$2.0.CO;2}.
\adsurl{2004JCli...17.4045B}.
\end{barticle}
\endbibitem

\bibitem[\protect\citeauthoryear{{Bhattacharyya} and
  {Okpala}}{2015}]{Bhattacharyya_2015}
\begin{barticle}
\bauthor{\bsnm{{Bhattacharyya}}, \binits{A.}},
\bauthor{\bsnm{{Okpala}}, \binits{K.C.}}:
\byear{2015},
\batitle{{Principal components of quiet time temporal variability of equatorial
  and low-latitude geomagnetic fields}}.
\bjtitle{J.\ Geophys.\ Res.}
\bvolume{120},
\bfpage{8799}.
\doiurl{https://doi.org/10.1002/2015JA021673}.
\end{barticle}
\endbibitem

\bibitem[\protect\citeauthoryear{{Booth} \textit{et~al.}}{2012}]{Booth_2012}
\begin{barticle}
\bauthor{\bsnm{{Booth}}, \binits{B.B.}},
\bauthor{\bsnm{{Dunstone}}, \binits{N.J.}},
\bauthor{\bsnm{{Halloran}}, \binits{P.R.}},
\bauthor{\bsnm{{Andrews}}, \binits{T.}},
\bauthor{\bsnm{{Bellouin}}, \binits{N.}}:
\byear{2012},
\batitle{{Aerosols implicated as a prime driver of twentieth-century North
  Atlantic climate variability}}.
\bjtitle{Nature}
\bvolume{484},
\bfpage{228–232}.
\doiurl{https://doi.org/10.1038/nature10946}.
\end{barticle}
\endbibitem

\bibitem[\protect\citeauthoryear{{Bro} and {Smilde}}{2014}]{Bro_2014}
\begin{barticle}
\bauthor{\bsnm{{Bro}}, \binits{R.}},
\bauthor{\bsnm{{Smilde}}, \binits{A.K.}}:
\byear{2014},
\batitle{{Principal component analysis}}.
\bjtitle{Anal. Methods}
\bvolume{6},
\bfpage{2812}.
\end{barticle}
\endbibitem

\bibitem[\protect\citeauthoryear{{Callendar}}{1938}]{Callendar_1938}
\begin{botherref}
\oauthor{\bsnm{{Callendar}}, \binits{G.S.}}:
1938,
{The artificial production of carbon dioxide and its influence on temperature}.
\textit{Quart.\ J.\ R.\ Met.\ Soc.},
223.
\doiurl{https://doi.org/10.1002/qj.49706427503}.
\end{botherref}
\endbibitem

\bibitem[\protect\citeauthoryear{{Cook} \textit{et~al.}}{2016}]{Cook_2016}
\begin{barticle}
\bauthor{\bsnm{{Cook}}, \binits{J.}},
\bauthor{\bsnm{{Oreskes}}, \binits{N.}},
\bauthor{\bsnm{{Doran}}, \binits{P.T.}},
\bauthor{\bsnm{{Anderegg}}, \binits{W.R.L.}},
\bauthor{\bsnm{{Verheggen}}, \binits{B.}},
\bauthor{\bsnm{{Maibach}}, \binits{E.W.}},
\bauthor{\bsnm{{Carlton}}, \binits{J.S.}},
\bauthor{\bsnm{{Lewandowsky}}, \binits{S.}},
\bauthor{\bsnm{{Skuce}}, \binits{A.G.}},
\bauthor{\bsnm{{Green}}, \binits{S.A.}},
\bauthor{\bsnm{{Nuccitelli}}, \binits{D.}},
\bauthor{\bsnm{{Jacobs}}, \binits{P.}},
\bauthor{\bsnm{{Richardson}}, \binits{M.}},
\bauthor{\bsnm{{Winkler}}, \binits{B.}},
\bauthor{\bsnm{{Painting}}, \binits{R.}},
\bauthor{\bsnm{{Rice}}, \binits{K.}}:
\byear{2016},
\batitle{{Consensus on consensus: a synthesis of consensus estimates on
  human-caused global warming}}.
\bjtitle{Environ. Res. Lett.}
\bvolume{11}(\bissue{4}),
\bfpage{048002}.
\doiurl{https://doi.org/10.1088/1748-9326/11/4/048002}.
\adsurl{2016ERL....11d8002C}.
\end{barticle}
\endbibitem

\bibitem[\protect\citeauthoryear{{Davy}, {Chen}, and {Hanna}}{2018}]{Davy_2018}
\begin{barticle}
\bauthor{\bsnm{{Davy}}, \binits{R.}},
\bauthor{\bsnm{{Chen}}, \binits{L.}},
\bauthor{\bsnm{{Hanna}}, \binits{E.}}:
\byear{2018},
\batitle{{Arctic amplification metrics}}.
\bjtitle{Int. J. Climatol.}
\bvolume{38},
\bfpage{4384}.
\doiurl{https://doi.org/10.1002/joc.5675}.
\end{barticle}
\endbibitem

\bibitem[\protect\citeauthoryear{{Dizerens}
  \textit{et~al.}}{2017}]{Dizerens_2017}
\begin{botherref}
\oauthor{\bsnm{{Dizerens}}, \binits{C.}},
\oauthor{\bsnm{{Lenggenhager}}, \binits{S.}},
\oauthor{\bsnm{{Schwander}}, \binits{A.} \bsuffix{M.~nad~{Buck}}},
\oauthor{\bsnm{{Foffa}}, \binits{S.}}:
2017,
{The 1956 Cold Wave inWestern Europe}.
\textit{Br{\o}nnimann, S. (Ed.) Historical Weather Extremes in Reanalyses.
  Geographica Bernensia G92},
101.
\doiurl{https://doi.org/10.4480/GB2017.G92.09}.
\end{botherref}
\endbibitem

\bibitem[\protect\citeauthoryear{{Egorova}
  \textit{et~al.}}{2018}]{Egorova_2018}
\begin{barticle}
\bauthor{\bsnm{{Egorova}}, \binits{T.}},
\bauthor{\bsnm{{Rozanov}}, \binits{E.}},
\bauthor{\bsnm{{Arsenovic}}, \binits{P.}},
\bauthor{\bsnm{{Peter}}, \binits{T.}},
\bauthor{\bsnm{{Schmutz}}, \binits{W.}}:
\byear{2018},
\batitle{{Contributions of Natural and Anthropogenic Forcing Agents to the
  Early 20th Century Warming}}.
\bjtitle{Front. Earth Sci.}
\bvolume{6}(\bissue{206}),
\bfpage{1}.
\doiurl{https://doi.org/10.3389/feart.2018.00206}.
\end{barticle}
\endbibitem

\bibitem[\protect\citeauthoryear{{Enfield}, {Mestas-Nu{\~n}ez}, and
  {Trimble}}{2001}]{Enfield_2001}
\begin{barticle}
\bauthor{\bsnm{{Enfield}}, \binits{D.B.}},
\bauthor{\bsnm{{Mestas-Nu{\~n}ez}}, \binits{A.M.}},
\bauthor{\bsnm{{Trimble}}, \binits{P.J.}}:
\byear{2001},
\batitle{{The Atlantic Multidecadal Oscillation and its relation to rainfall
  and river flows in the continental U.S.}}
\bjtitle{Geophys.\ Res.\ Lett.}
\bvolume{28}(\bissue{10}),
\bfpage{2077}.
\doiurl{https://doi.org/10.1029/2000GL012745}.
\end{barticle}
\endbibitem

\bibitem[\protect\citeauthoryear{{Frajka-Williams}, {Beaulieu}, and
  {Duchez}}{2017}]{Frajka-Williams_2017}
\begin{barticle}
\bauthor{\bsnm{{Frajka-Williams}}, \binits{E.}},
\bauthor{\bsnm{{Beaulieu}}, \binits{C.}},
\bauthor{\bsnm{{Duchez}}, \binits{A.}}:
\byear{2017},
\batitle{{Emerging negative Atlantic Multidecadal Oscillation index in spite of
  warm subtropics}}.
\bjtitle{Scientific Reports}
\bvolume{7},
\bfpage{11224}.
\doiurl{https://doi.org/10.1038/s41598-017-11046-x}.
\adsurl{2017NatSR...711224F}.
\end{barticle}
\endbibitem

\bibitem[\protect\citeauthoryear{{Friedman}
  \textit{et~al.}}{2013}]{Friedman_2013}
\begin{barticle}
\bauthor{\bsnm{{Friedman}}, \binits{A.R.}},
\bauthor{\bsnm{{Hwang}}, \binits{Y.-T.}},
\bauthor{\bsnm{{Chiang}}, \binits{J.C.H.}},
\bauthor{\bsnm{{Frierson}}, \binits{D.M.W.}}:
\byear{2013},
\batitle{{Interhemispheric Temperature Asymmetry over the Twentieth Century and
  in Future Projections}}.
\bjtitle{J. Clim.}
\bvolume{26},
\bfpage{5419}.
\doiurl{https://doi.org/10.1175/JCLI-D-12-00525.1}.
\adsurl{2013JCli...26.5419F}.
\end{barticle}
\endbibitem

\bibitem[\protect\citeauthoryear{{Fu} \textit{et~al.}}{1999}]{Fu_1999}
\begin{barticle}
\bauthor{\bsnm{{Fu}}, \binits{C.}},
\bauthor{\bsnm{{Diaz}}, \binits{H.F.}},
\bauthor{\bsnm{{Dong}}, \binits{D.}},
\bauthor{\bsnm{{Fletcher}}, \binits{J.O.}}:
\byear{1999},
\batitle{{Changes in atmospheric circulation over northern hemisphere oceans
  associated with the rapid warming of the 1920s}}.
\bjtitle{Int. J. Climatol.}
\bvolume{19},
\bfpage{581}.
\doiurl{https://doi.org/10.1002/(SICI)1097-0088(199905)19:6$<$581::AID-JOC396$>$3.3.CO;2-G}.
\adsurl{1999IJCli..19..581F}.
\end{barticle}
\endbibitem

\bibitem[\protect\citeauthoryear{{Hannachi}, {Jolliffe}, and
  {Stephenson}}{2007}]{Hannachi_2007}
\begin{barticle}
\bauthor{\bsnm{{Hannachi}}, \binits{A.}},
\bauthor{\bsnm{{Jolliffe}}, \binits{I.T.}},
\bauthor{\bsnm{{Stephenson}}, \binits{D.B.}}:
\byear{2007},
\batitle{{Empirical orthogonal functions and related techniques in atmospheric
  science: A review}}.
\bjtitle{Int.\ J.\ Clim.}
\bvolume{27},
\bfpage{1119}.
\doiurl{https://doi.org/10.1002/joc.1499}.
\end{barticle}
\endbibitem

\bibitem[\protect\citeauthoryear{{Hansen} \textit{et~al.}}{1981}]{Hansen_1981}
\begin{barticle}
\bauthor{\bsnm{{Hansen}}, \binits{J.}},
\bauthor{\bsnm{{Johnson}}, \binits{D.}},
\bauthor{\bsnm{{Lacis}}, \binits{A.}},
\bauthor{\bsnm{{Lebedeff}}, \binits{S.}},
\bauthor{\bsnm{{Lee}}, \binits{P.}},
\bauthor{\bsnm{{Rind}}, \binits{D.}},
\bauthor{\bsnm{{Russell}}, \binits{G.}}:
\byear{1981},
\batitle{{Climate Impact of Increasing Atmospheric Carbon Dioxide}}.
\bjtitle{Science}
\bvolume{213},
\bfpage{957}.
\doiurl{https://doi.org/10.1126/science.213.4511.957}.
\end{barticle}
\endbibitem

\bibitem[\protect\citeauthoryear{{Harris}}{2010}]{Harris_2010}
\begin{barticle}
\bauthor{\bsnm{{Harris}}, \binits{D.C.}}:
\byear{2010},
\batitle{{Charles David Keeling and the story of atmospheric CO2
  measurements}}.
\bjtitle{Anal. Chem.}
\bvolume{82},
\bfpage{7865–70}.
\doiurl{https://doi.org/10.1021/ac1001492}.
\end{barticle}
\endbibitem

\bibitem[\protect\citeauthoryear{{Hawkins} and {Jones}}{2013}]{Hawkins_2013}
\begin{barticle}
\bauthor{\bsnm{{Hawkins}}, \binits{E.}},
\bauthor{\bsnm{{Jones}}, \binits{P.D.}}:
\byear{2013},
\batitle{{On increasing global temperatures: 75 years after Callendar}}.
\bjtitle{Quart.\ J.\ R.\ Met.\ Soc.}
\bvolume{139},
\bfpage{1961}.
\doiurl{https://doi.org/10.1002/qj.2178}.
\adsurl{2013QJRMS.139.1961H}.
\end{barticle}
\endbibitem

\bibitem[\protect\citeauthoryear{{Haywood} and {Boucher}}{2000}]{Haywood_2000}
\begin{barticle}
\bauthor{\bsnm{{Haywood}}, \binits{J.}},
\bauthor{\bsnm{{Boucher}}, \binits{O.}}:
\byear{2000},
\batitle{{Estimates of the direct and indirect radiative forcing due to
  tropospheric aerosols: A review}}.
\bjtitle{Rev. Geophys.}
\bvolume{38},
\bfpage{513}.
\doiurl{https://doi.org/10.1029/1999RG000078}.
\end{barticle}
\endbibitem

\bibitem[\protect\citeauthoryear{{Hegerl} \textit{et~al.}}{2018}]{Hegerl_2018}
\begin{botherref}
\oauthor{\bsnm{{Hegerl}}, \binits{G.C.}},
\oauthor{\bsnm{{Br{\"o}nnimann}}, \binits{S.}},
\oauthor{\bsnm{{Schurer}}, \binits{A.}},
\oauthor{\bsnm{{Cowan}}, \binits{T.}}:
2018,
{The early 20th century warming: Anomalies, causes and consequences}.
\textit{WIREs Clim. Change}
\textbf{9}(e522).
\url{https://doi.org/10.1002/wcc.522}.
\end{botherref}
\endbibitem

\bibitem[\protect\citeauthoryear{{Hodson}, {Robson}, and
  {Sutton}}{2014}]{Hodson_2014}
\begin{barticle}
\bauthor{\bsnm{{Hodson}}, \binits{D.L.R.}},
\bauthor{\bsnm{{Robson}}, \binits{J.I.}},
\bauthor{\bsnm{{Sutton}}, \binits{R.T.}}:
\byear{2014},
\batitle{{An Anatomy of the Cooling of the North Atlantic Ocean in the 1960s
  and 1970s}}.
\bjtitle{J. Clim.}
\bvolume{27},
\bfpage{8229}.
\doiurl{https://doi.org/10.1175/JCLI-D-14-00301.1}.
\adsurl{2014JCli...27.8229H}.
\end{barticle}
\endbibitem

\bibitem[\protect\citeauthoryear{{Holappa}
  \textit{et~al.}}{2014}]{Holappa_2014}
\begin{barticle}
\bauthor{\bsnm{{Holappa}}, \binits{L.}},
\bauthor{\bsnm{{Mursula}}, \binits{K.}},
\bauthor{\bsnm{{Asikainen}}, \binits{T.}},
\bauthor{\bsnm{{Richardson}}, \binits{I.G.}}:
\byear{2014},
\batitle{{Annual fractions of high-speed streams from principal component
  analysis of local geomagnetic activity}}.
\bjtitle{J.\ Geophys.\ Res.}
\bvolume{119},
\bfpage{4544}.
\doiurl{https://doi.org/10.1002/2014JA019958}.
\adsurl{2014JGRA..119.4544H}.
\end{barticle}
\endbibitem

\bibitem[\protect\citeauthoryear{{Johannessen}
  \textit{et~al.}}{2016}]{Johannessen_2016}
\begin{barticle}
\bauthor{\bsnm{{Johannessen}}, \binits{O.M.}},
\bauthor{\bsnm{{Kuzmina}}, \binits{S.I.}},
\bauthor{\bsnm{{Bobylev}}, \binits{L.P.}},
\bauthor{\bsnm{{Miles}}, \binits{M.W.}}:
\byear{2016},
\batitle{{Surface air temperature variability and trends in the Arctic: new
  amplification assessment and regionalisation}}.
\bjtitle{Tellus A}
\bvolume{68},
\bfpage{28234}.
\doiurl{https://doi.org/10.3402/tellusa.v68.28234}.
\adsurl{2016TellA..6828234J}.
\end{barticle}
\endbibitem

\bibitem[\protect\citeauthoryear{{Kendall} and {Stuart}}{1967}]{Kendall_1967}
\begin{bbook}
\bauthor{\bsnm{{Kendall}}, \binits{M.G.}},
\bauthor{\bsnm{{Stuart}}, \binits{A.}}:
\byear{1967},
\bbtitle{{The advanced theory of statistics. Vol.2: Inference and
  relationship}},
\bpublisher{London: Griffin, 1979, 2th ed.}, \blocation{???}.
\end{bbook}
\endbibitem

\bibitem[\protect\citeauthoryear{{Killick}, {Fearnhead}, and
  {Eckley}}{2012}]{Killick_2012}
\begin{botherref}
\oauthor{\bsnm{{Killick}}, \binits{R.}},
\oauthor{\bsnm{{Fearnhead}}, \binits{P.}},
\oauthor{\bsnm{{Eckley}}, \binits{I.A.}}:
2012,
{Optimal detection of changepoints with a linear computational cost}.
\textit{Journal of American Statistical Association},
1590.
\end{botherref}
\endbibitem

\bibitem[\protect\citeauthoryear{{Kumar}, {Rai}, and
  {Kumar}}{2008}]{Kumar_2008}
\begin{barticle}
\bauthor{\bsnm{{Kumar}}, \binits{D.}},
\bauthor{\bsnm{{Rai}}, \binits{C.S.}},
\bauthor{\bsnm{{Kumar}}, \binits{S.}}:
\byear{2008},
\batitle{{Principal Component Analysis for Data Compression and Face
  Recognition}}.
\bjtitle{INFOCOMP}
\bvolume{7},
\bfpage{48}.
\end{barticle}
\endbibitem

\bibitem[\protect\citeauthoryear{{Lavielle}}{2005}]{Lavielle_2005}
\begin{barticle}
\bauthor{\bsnm{{Lavielle}}, \binits{M.}}:
\byear{2005},
\batitle{{Using penalized contrasts for the change-point problem}}.
\bjtitle{Sinal Processing}
\bvolume{85},
\bfpage{1501}.
\end{barticle}
\endbibitem

\bibitem[\protect\citeauthoryear{{Lenssen}
  \textit{et~al.}}{2019}]{Lenssen_2019}
\begin{barticle}
\bauthor{\bsnm{{Lenssen}}, \binits{N.J.L.}},
\bauthor{\bsnm{{Schmidt}}, \binits{G.A.}},
\bauthor{\bsnm{{Hansen}}, \binits{J.E.}},
\bauthor{\bsnm{{Menne}}, \binits{M.J.}},
\bauthor{\bsnm{{Persin}}, \binits{A.}},
\bauthor{\bsnm{{Ruedy}}, \binits{R.}},
\bauthor{\bsnm{{Zyss}}, \binits{D.}}:
\byear{2019},
\batitle{{Improvements in the GISTEMP Uncertainty Model}}.
\bjtitle{J.\ Geophys.\ Res.}
\bvolume{124}(\bissue{12}),
\bfpage{6307}.
\doiurl{https://doi.org/10.1029/2018JD029522}.
\adsurl{2019JGRD..124.6307L}.
\end{barticle}
\endbibitem

\bibitem[\protect\citeauthoryear{{Mann}, {Bradley}, and
  {Hughes}}{1998}]{Mann_1998}
\begin{barticle}
\bauthor{\bsnm{{Mann}}, \binits{M.E.}},
\bauthor{\bsnm{{Bradley}}, \binits{R.S.}},
\bauthor{\bsnm{{Hughes}}, \binits{M.K.}}:
\byear{1998},
\batitle{{Global-scale temperature patterns and climate forcing over the past
  six centuries}}.
\bjtitle{Nature}
\bvolume{392},
\bfpage{779}.
\doiurl{https://doi.org/10.1038/33859}.
\adsurl{1998Natur.392..779M}.
\end{barticle}
\endbibitem

\bibitem[\protect\citeauthoryear{{Mantua} and {Hare}}{2002}]{Mantua_2002}
\begin{barticle}
\bauthor{\bsnm{{Mantua}}, \binits{N.J.}},
\bauthor{\bsnm{{Hare}}, \binits{S.}}:
\byear{2002},
\batitle{{The Pacific Decadal Oscillation}}.
\bjtitle{J. Oceanogr.}
\bvolume{58},
\bfpage{35}.
\doiurl{https://doi.org/10.1023/A:1015820616384}.
\end{barticle}
\endbibitem

\bibitem[\protect\citeauthoryear{{Meehl} \textit{et~al.}}{2003}]{Meehl_2003}
\begin{barticle}
\bauthor{\bsnm{{Meehl}}, \binits{G.A.}},
\bauthor{\bsnm{{Washington}}, \binits{W.M.}},
\bauthor{\bsnm{{Wigley}}, \binits{T.M.L.}},
\bauthor{\bsnm{{Arblaster}}, \binits{J.M.}},
\bauthor{\bsnm{{Dai}}, \binits{A.}}:
\byear{2003},
\batitle{{Solar and Greenhouse Gas Forcing and Climate Response in the
  Twentieth Century.}}
\bjtitle{J. Clim.}
\bvolume{16},
\bfpage{426}.
\doiurl{https://doi.org/10.1175/1520-0442(2003)016<0426:SAGGFA>2.0.CO;2}.
\adsurl{2003JCli...16..426M}.
\end{barticle}
\endbibitem

\bibitem[\protect\citeauthoryear{{Menne} \textit{et~al.}}{2018}]{Menne_2018}
\begin{barticle}
\bauthor{\bsnm{{Menne}}, \binits{M.J.}},
\bauthor{\bsnm{{Williams}}, \binits{C.N.}},
\bauthor{\bsnm{{Gleason}}, \binits{B.E.}},
\bauthor{\bsnm{{Rennie}}, \binits{J.J.}},
\bauthor{\bsnm{{Lawrimore}}, \binits{J.H.}}:
\byear{2018},
\batitle{{The Global Historical Climatology Network Monthly Temperature
  Dataset, Version 4}}.
\bjtitle{Journal of Climate}
\bvolume{31}(\bissue{24}),
\bfpage{9835}.
\doiurl{https://doi.org/10.1175/JCLI-D-18-0094.1}.
\adsurl{2018JCli...31.9835M}.
\end{barticle}
\endbibitem

\bibitem[\protect\citeauthoryear{{Myhre}}{2009}]{Myhre_2009}
\begin{barticle}
\bauthor{\bsnm{{Myhre}}, \binits{G.}}:
\byear{2009},
\batitle{{Consistency Between Satellite-Derived and Modeled Estimates of the
  Direct Aerosol Effect}}.
\bjtitle{Science}
\bvolume{325},
\bfpage{187}.
\doiurl{https://doi.org/10.1126/science.1174461}.
\adsurl{2009Sci...325..187M}.
\end{barticle}
\endbibitem

\bibitem[\protect\citeauthoryear{{Newman} \textit{et~al.}}{2016}]{Newman_2016}
\begin{barticle}
\bauthor{\bsnm{{Newman}}, \binits{M.}},
\bauthor{\bsnm{{Alexander}}, \binits{M.A.}},
\bauthor{\bsnm{{Ault}}, \binits{T.R.}},
\bauthor{\bsnm{{Cobb}}, \binits{K.M.}},
\bauthor{\bsnm{{Deser}}, \binits{C.}},
\bauthor{\bsnm{{Di Lorenzo}}, \binits{E.}},
\bauthor{\bsnm{{Mantua}}, \binits{N.J.}},
\bauthor{\bsnm{{Miller}}, \binits{A.J.}},
\bauthor{\bsnm{{Minobe}}, \binits{S.}},
\bauthor{\bsnm{{Nakamura}}, \binits{H.}},
\bauthor{\bsnm{{Schneider}}, \binits{N.}},
\bauthor{\bsnm{{Vimont}}, \binits{D.J.}},
\bauthor{\bsnm{{Phillips}}, \binits{A.S.}},
\bauthor{\bsnm{{Scott}}, \binits{J.D.}},
\bauthor{\bsnm{{Smith}}, \binits{C.A.}}:
\byear{2016},
\batitle{{The Pacific Decadal Oscillation, Revisited}}.
\bjtitle{J. Clim.}
\bvolume{29},
\bfpage{4399}.
\doiurl{https://doi.org/10.1175/JCLI-D-15-0508.1}.
\adsurl{2016JCli...29.4399N}.
\end{barticle}
\endbibitem

\bibitem[\protect\citeauthoryear{{Nicholls}}{2007}]{Nicholls_2007}
\begin{barticle}
\bauthor{\bsnm{{Nicholls}}, \binits{N.}}:
\byear{2007},
\batitle{{Man-made Carbon Dioxide and the ``Greenhouse'' Effect}}.
\bjtitle{Nature}
\bvolume{448},
\bfpage{992}.
\end{barticle}
\endbibitem

\bibitem[\protect\citeauthoryear{{Nozawa} \textit{et~al.}}{2005}]{Nozawa_2005}
\begin{barticle}
\bauthor{\bsnm{{Nozawa}}, \binits{T.}},
\bauthor{\bsnm{{Nagashima}}, \binits{T.}},
\bauthor{\bsnm{{Shiogama}}, \binits{H.}},
\bauthor{\bsnm{{Crooks}}, \binits{S.A.}}:
\byear{2005},
\batitle{{Detecting natural influence on surface air temperature change in the
  early twentieth century}}.
\bjtitle{Geophys.\ Res.\ Lett.}
\bvolume{32},
\bfpage{L20719}.
\doiurl{https://doi.org/10.1029/2005GL023540}.
\end{barticle}
\endbibitem

\bibitem[\protect\citeauthoryear{{Ohmura}}{2009}]{Ohmura_2009}
\begin{barticle}
\bauthor{\bsnm{{Ohmura}}, \binits{A.}}:
\byear{2009},
\batitle{{Observed decadal variations in surface solar radiation and their
  causes}}.
\bjtitle{J.\ Geophys.\ Res.}
\bvolume{114},
\bfpage{D00D05}.
\doiurl{https://doi.org/10.1029/2008JD011290}.
\end{barticle}
\endbibitem

\bibitem[\protect\citeauthoryear{{Polyakov}
  \textit{et~al.}}{2010}]{Polyakov_2010}
\begin{barticle}
\bauthor{\bsnm{{Polyakov}}, \binits{I.V.}},
\bauthor{\bsnm{{Alexeev}}, \binits{V.A.}},
\bauthor{\bsnm{{Bhatt}}, \binits{U.S.}},
\bauthor{\bsnm{{Polyakova}}, \binits{E.I.}},
\bauthor{\bsnm{{Zhang}}, \binits{X.}}:
\byear{2010},
\batitle{{North Atlantic warming: patterns of long-term trend and multidecadal
  variability}}.
\bjtitle{Clim. Dynam.}
\bvolume{34},
\bfpage{459}.
\doiurl{https://doi.org/10.1007/s00382-009-0589-5}.
\adsurl{2010ClDy...34..459P}.
\end{barticle}
\endbibitem

\bibitem[\protect\citeauthoryear{{Reid}}{2013}]{Reid_1997}
\begin{barticle}
\bauthor{\bsnm{{Reid}}, \binits{G.C.}}:
\byear{2013},
\batitle{{Solar Forcing of Global Climate Change Since The Mid-17th Century}}.
\bjtitle{Climatic Change}
\bvolume{37},
\bfpage{391}.
\end{barticle}
\endbibitem

\bibitem[\protect\citeauthoryear{{Revelle} and {Suess}}{1957}]{Revelle_1957}
\begin{barticle}
\bauthor{\bsnm{{Revelle}}, \binits{R.}},
\bauthor{\bsnm{{Suess}}, \binits{H.E.}}:
\byear{1957},
\batitle{{Carbon Dioxide Exchange between Atmosphere and Ocean, and the
  Question of an Increase of Atmospheric CO$_{2 }$during the Past Decades}}.
\bjtitle{Tellus}
\bvolume{9},
\bfpage{18}.
\adsurl{1957Tell....9...18R}.
\end{barticle}
\endbibitem

\bibitem[\protect\citeauthoryear{{Robock} and {Free}}{1995}]{Robock_1995}
\begin{barticle}
\bauthor{\bsnm{{Robock}}, \binits{A.}},
\bauthor{\bsnm{{Free}}, \binits{M.P.}}:
\byear{1995},
\batitle{{Ice cores as an index of global volcanism from 1850 to the present}}.
\bjtitle{J.\ Geophys.\ Res.}
\bvolume{100},
\bfpage{11549}.
\doiurl{https://doi.org/10.1029/95JD00825}.
\end{barticle}
\endbibitem

\bibitem[\protect\citeauthoryear{{Robson}, {Sutton}, and
  {Smith}}{2014}]{Robson_2014}
\begin{bchapter}
\bauthor{\bsnm{{Robson}}, \binits{J.}},
\bauthor{\bsnm{{Sutton}}, \binits{R.}},
\bauthor{\bsnm{{Smith}}, \binits{D.}}:
\byear{2014},
\bctitle{{Decadal predictions of the cooling and freshening of the North
  Atlantic in the 1960s and the role of ocean circulation}}.
In: \bbtitle{EGU General Assembly Conference Abstracts},
\bsertitle{EGU General Assembly Conference Abstracts}
\bseriesno{16},
\bfpage{14052}.
\adsurl{2014EGUGA..1614052R}.
\end{bchapter}
\endbibitem

\bibitem[\protect\citeauthoryear{{Rybski} \textit{et~al.}}{2006}]{Rybski_2006}
\begin{barticle}
\bauthor{\bsnm{{Rybski}}, \binits{D.}},
\bauthor{\bsnm{{Bunde}}, \binits{A.}},
\bauthor{\bsnm{{Havlin}}, \binits{S.}},
\bauthor{\bsnm{{von Storch}}, \binits{H.}}:
\byear{2006},
\batitle{{Long-term persistence in climate and the detection problem}}.
\bjtitle{Geophys.\ Res.\ Lett.}
\bvolume{33},
\bfpage{L06718}.
\doiurl{https://doi.org/10.1029/2005GL025591}.
\adsurl{2006GeoRL..33.6718R}.
\end{barticle}
\endbibitem

\bibitem[\protect\citeauthoryear{{Salinger} and {Gunn}}{1975}]{Salinger_1975}
\begin{barticle}
\bauthor{\bsnm{{Salinger}}, \binits{M.J.}},
\bauthor{\bsnm{{Gunn}}, \binits{J.M.}}:
\byear{1975},
\batitle{{Recent climatic warming around New Zealand}}.
\bjtitle{Nature}
\bvolume{256},
\bfpage{396}.
\doiurl{https://doi.org/10.1038/256396a0}.
\end{barticle}
\endbibitem

\bibitem[\protect\citeauthoryear{{Salinger}, {Renwick}, and
  {Mullan}}{2001}]{Salinger_2001}
\begin{barticle}
\bauthor{\bsnm{{Salinger}}, \binits{M.J.}},
\bauthor{\bsnm{{Renwick}}, \binits{J.A.}},
\bauthor{\bsnm{{Mullan}}, \binits{A.B.}}:
\byear{2001},
\batitle{{Interdecadal Pacific Oscillation and South Pacific climate}}.
\bjtitle{Int. J. Climatol.}
\bvolume{21},
\bfpage{1705}.
\doiurl{https://doi.org/10.1002/joc.691}.
\adsurl{2001IJCli..21.1705S}.
\end{barticle}
\endbibitem

\bibitem[\protect\citeauthoryear{{Sawyer}}{1972}]{Sawyer_1972}
\begin{barticle}
\bauthor{\bsnm{{Sawyer}}, \binits{J.S.}}:
\byear{1972},
\batitle{{Man-made Carbon Dioxide and the ``Greenhouse'' Effect}}.
\bjtitle{Nature}
\bvolume{239},
\bfpage{23}.
\doiurl{https://doi.org/10.1038/239023a0}.
\adsurl{1972Natur.239...23S}.
\end{barticle}
\endbibitem

\bibitem[\protect\citeauthoryear{{Schlesinger} and
  {Ramankutty}}{1994}]{Schlesinger_1994}
\begin{barticle}
\bauthor{\bsnm{{Schlesinger}}, \binits{M.E.}},
\bauthor{\bsnm{{Ramankutty}}, \binits{N.}}:
\byear{1994},
\batitle{{An oscillation in the global climate system of period 65-70 years}}.
\bjtitle{Nature}
\bvolume{367},
\bfpage{723}.
\doiurl{https://doi.org/10.1038/367723a0}.
\adsurl{1994Natur.367..723S}.
\end{barticle}
\endbibitem

\bibitem[\protect\citeauthoryear{{Shapiro}
  \textit{et~al.}}{2011}]{Shapiro_2011}
\begin{barticle}
\bauthor{\bsnm{{Shapiro}}, \binits{A.I.}},
\bauthor{\bsnm{{Schmutz}}, \binits{W.}},
\bauthor{\bsnm{{Rozanov}}, \binits{E.}},
\bauthor{\bsnm{{Schoell}}, \binits{M.}},
\bauthor{\bsnm{{Haberreiter}}, \binits{M.}},
\bauthor{\bsnm{{Shapiro}}, \binits{A.V.}},
\bauthor{\bsnm{{Nyeki}}, \binits{S.}}:
\byear{2011},
\batitle{{A new approach to the long-term reconstruction of the solar
  irradiance leads to large historical solar forcing}}.
\bjtitle{Astron. Astrophys.}
\bvolume{529},
\bfpage{A67}.
\doiurl{https://doi.org/10.1051/0004-6361/201016173}.
\adsurl{2011A\%26A...529A..67S}.
\end{barticle}
\endbibitem

\bibitem[\protect\citeauthoryear{{Siegel}}{1988}]{Siegel_1988}
\begin{bbook}
\bauthor{\bsnm{{Siegel}}, \binits{N.J.} \bsuffix{S.and~{Castellan}}}:
\byear{1988},
\bbtitle{{Nonparametric statistics for the behavioral sciences}},
\bpublisher{International Edition - McGraw-Hill Book Co., Singapore},
  \blocation{???},
\bfpage{58}.
\adsurl{1956nsbs.book.....S}.
\end{bbook}
\endbibitem

\bibitem[\protect\citeauthoryear{{Suo} \textit{et~al.}}{2013}]{Suo_2013}
\begin{barticle}
\bauthor{\bsnm{{Suo}}, \binits{L.}},
\bauthor{\bsnm{{Otter{\aa}}}, \binits{O.H.}},
\bauthor{\bsnm{{Bentsen}}, \binits{M.}},
\bauthor{\bsnm{{Gao}}, \binits{Y.}},
\bauthor{\bsnm{{Johannessen}}, \binits{O.M.}}:
\byear{2013},
\batitle{{External forcing of the early 20th century Arctic warming}}.
\bjtitle{Tellus A}
\bvolume{65},
\bfpage{20578}.
\doiurl{https://doi.org/10.3402/tellusa.v65i0.20578}.
\adsurl{2013TellA..6520578S}.
\end{barticle}
\endbibitem

\bibitem[\protect\citeauthoryear{{Takalo} and {Mursula}}{2018}]{Takalo_2018}
\begin{barticle}
\bauthor{\bsnm{{Takalo}}, \binits{J.}},
\bauthor{\bsnm{{Mursula}}, \binits{K.}}:
\byear{2018},
\batitle{{Principal component analysis of sunspot cycle shape}}.
\bjtitle{Astron. Astrophys.}
\bvolume{620},
\bfpage{100}.
\doiurl{https://doi.org/10.1051/0004-6361/201833924}.
\end{barticle}
\endbibitem

\bibitem[\protect\citeauthoryear{{Trenberth} and {Shea}}{2006}]{Trenberth_2006}
\begin{barticle}
\bauthor{\bsnm{{Trenberth}}, \binits{K.E.}},
\bauthor{\bsnm{{Shea}}, \binits{D.J.}}:
\byear{2006},
\batitle{{Atlantic hurricanes and natural variability in 2005}}.
\bjtitle{Geophys.\ Res.\ Lett.}
\bvolume{33}(\bissue{12}),
\bfpage{L12704}.
\doiurl{https://doi.org/10.1029/2006GL026894}.
\adsurl{2006GeoRL..3312704T}.
\end{barticle}
\endbibitem

\bibitem[\protect\citeauthoryear{{Twardosz} and
  {Kossowska-Cezak}}{2016}]{Twardosz_2016}
\begin{barticle}
\bauthor{\bsnm{{Twardosz}}, \binits{R.}},
\bauthor{\bsnm{{Kossowska-Cezak}}, \binits{U.}}:
\byear{2016},
\batitle{{Exceptionally cold and mild winters in Europe (1951-2010)}}.
\bjtitle{Theoretical and Applied Climatology}
\bvolume{125}(\bissue{1-2}),
\bfpage{399}.
\doiurl{https://doi.org/10.1007/s00704-015-1524-9}.
\adsurl{2016ThApC.125..399T}.
\end{barticle}
\endbibitem

\bibitem[\protect\citeauthoryear{{Willmott} and
  {Matsuura}}{2018}]{Willmott_2018}
\begin{botherref}
\oauthor{\bsnm{{Willmott}}, \binits{C.J.}},
\oauthor{\bsnm{{Matsuura}}, \binits{K.}}:
2018,
\textit{{Terrestrial Air Temperature and Precipitation: Monthly and Annual Time
  Series (1900-2017)}}.
\url{http://climate.geog.udel.edu/$~$climate/htmlpages/Global2017/
  README.GlobalTsT2017.html}.
\end{botherref}
\endbibitem

\bibitem[\protect\citeauthoryear{{Winchester}}{2017}]{Winchester_2017}
\begin{botherref}
\oauthor{\bsnm{{Winchester}}, \binits{S.}}:
2017,
{How the Pacific Ocean changes the weather around the world}.
\textit{Popular Science}.
\end{botherref}
\endbibitem

\bibitem[\protect\citeauthoryear{{Wood} \textit{et~al.}}{2010}]{Wood_2010}
\begin{botherref}
\oauthor{\bsnm{{Wood}}, \binits{K.R.}},
\oauthor{\bsnm{{Overland}}, \binits{J.E.}},
\oauthor{\bsnm{{J{\'o}nsson}}, \binits{T.}},
\oauthor{\bsnm{{Smoliak}}, \binits{B.V.}}:
2010,
{Air temperature variations on the Atlantic - Arctic boundary since 1802: the
  low-frequency pattern and ocean teleconnections}.
\textit{AGU Fall Meeting Abstracts},
GC44A.
\adsurl{2010AGUFMGC44A..07W}.
\end{botherref}
\endbibitem

\bibitem[\protect\citeauthoryear{{Yamanouchi}}{2011}]{Yamanouchi_2011}
\begin{barticle}
\bauthor{\bsnm{{Yamanouchi}}, \binits{T.}}:
\byear{2011},
\batitle{{Early 20th century warming in the Arctic: A review}}.
\bjtitle{Polar Sci.}
\bvolume{5},
\bfpage{53}.
\doiurl{https://doi.org/10.1016/j.polar.2010.10.002}.
\adsurl{2011PolSc...5...53Y}.
\end{barticle}
\endbibitem

\end{thebibliography}
%
%

\end{document}